\documentclass{article}

\usepackage{PRIMEarxiv}

\usepackage{moreverb,url}
\usepackage{bm}
\usepackage{graphicx}
\usepackage{soul}
\usepackage{xcolor}
\usepackage{prodint}
\usepackage{caption, subcaption}
\usepackage{amsmath}
\usepackage{amsthm}
\usepackage{mathtools}
\usepackage{dsfont}
\usepackage{multirow}
\usepackage{makecell}
\usepackage[symbol, bottom]{footmisc}

\newtheorem{proposition}{Proposition}

\newcommand{\given}{\, | \,}

\usepackage{tikz}
\usetikzlibrary{arrows}

\usepackage{algorithm}
\usepackage{algpseudocode}

\usepackage[utf8]{inputenc} 
\usepackage[T1]{fontenc}    
\usepackage{hyperref}       
\usepackage{url}            
\usepackage{booktabs}       
\usepackage{amsfonts}       
\usepackage{nicefrac}       
\usepackage{microtype}      
\usepackage{lipsum}
\usepackage{fancyhdr}       
\usepackage{graphicx}       
\graphicspath{{figures/}}     

\title{Non-parametric estimation of transition intensities in interval censored Markov multi-state models without loops
}

\author{
}

\begin{document}
\maketitle

\vspace{-7em}
\begin{center}
\large{
Daniel Gomon \textsuperscript{1,}\footnote[1]{Corresponding Author, d.gomon@math.leidenuniv.nl},
Hein Putter \textsuperscript{1,2}}
\\
\bigskip
\normalsize
\textit{\textsuperscript{1} Mathematical Institute, Leiden University, Einsteinweg 55, Leiden, the Netherlands}
\\
\textit{\textsuperscript{2} Department of Biomedical Data Sciences, Leiden University Medical Centre, Einthovenweg 20, Leiden, the Netherlands}
\\
\bigskip
\end{center}

\begin{abstract}

Interval-censored multi state data is collected when the state of a subject is observed periodically. The analysis of such data using non-parametric multi-state models was not possible until recently, but is very desirable as it allows for more flexibility than its parametric counterparts. The single available result to date has some unique drawbacks. We propose a non-parametric estimator of the transition intensities for interval-censored multi state data using an Expectation Maximisation algorithm. The method allows for a mix of interval-censored and right-censored (exactly observed) transitions. A condition to check for the convergence of the algorithm is given. A simulation study comparing the proposed estimator to a consistent estimator is performed, and shown to yield near identical estimates at smaller computational cost. A data set on the emergence of teeth in children is analysed. Software to perform the analyses is publicly available.
\end{abstract}

\fancyhead[LO]{Gomon and Putter}


\section{Introduction}
\label{sec:introduction}

Non-parametric estimators such as the Kaplan-Meier survival curve are very popular in the standard survival setting with right-censored time to event data. Their popularity comes from their flexibility, as they do not require assumptions to be made on the underlying form of the hazard rate. Research in survival analysis is very focused on right-censored data, where event times are either observed exactly or known to occur after a certain censoring time. Many real life scenarios however generate interval-censored data, where event times are only known to lie between two observation times. Such data often arises in medical studies, where patients are observed periodically over an extended interval of time.  Non-parametric estimators have also been derived for interval-censored data. Examples and an overview of analysis techniques for interval-censored data are given in Bogaerts et al. \cite{Bogaerts2020} An important quantity in this setting is Turnbull's non-parametric estimator for the survival function.\cite{Turnbull1976} Due to the additional complexity of interval-censored data, the estimator requires the use of an Expectation Maximization (EM) algorithm. It might therefore be desirable to negate the need to use interval-censored techniques, for example by using imputation techniques to recover the ``missing'' event time or ignore the missingness mechanism altogether and using techniques for right-censored data instead. These approaches can lead to biases and incorrect inference.\cite{Lindsey1998, Commenges2002} It is therefore important to use appropriate techniques in the presence of interval-censored data.

\begin{figure}[!ht]
\centering
    \includegraphics[width = 0.45\textwidth]{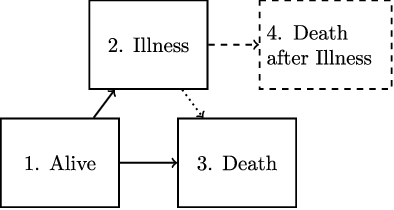}
    \caption{Graphical display of the (extended) illness-death model. Extended: solid and dashed lines. Standard: solid and dotted lines.}
    \label{fig:MSmodelsgraphical}
\end{figure}

The standard survival setting with a single time-to-event outcome is often insufficient to describe the data-generating mechanism of a study. As an example, illness often precedes death in medical studies and the interest may lie in modelling both the time to death and time to illness. This situation can be well described using an (extended) illness-death model (Figure \ref{fig:MSmodelsgraphical}), where the extended illness-death model allows to distinguish between initially healthy subjects that have experienced illness and those who have not when calculating probabilities of experiencing death. These models are examples of a multi-state model, a powerful generalisation of the standard survival setting where the data is described using states (such as illness, death) and transitions (becoming ill, dying when healthy). Under the Markov assumption, the cumulative intensities can be non-parametrically estimated for multi-state models with right-censored data using the Nelson-Aalen estimator and translated into transition probabilities using the Aalen-Johanssen estimator.\cite{AalenJohansen1978} Theory and practical applications are well described in many textbooks and tutorials.\cite{Cook2020, Andersen_Ravn_2024, Putter2006}

In a multi-state setting, data is also often interval-censored when a subject's state is observed periodically. One setting which leads to such interval-censored multi state data originates from studies based on \textit{panel data}. Many demographic studies \cite{SHARE, AHEAD} accrue one of more cohorts of subjects that are periodically assessed, often using questionnaires. Frequently these assessments (waves) are very regular in calendar time, but due to the time scale used in the analysis (age of subjects) such data typically leads to subject-specific assessment times.  

Contrary to the right-censored setting, there are very few results that allow for the estimation of (cumulative) intensity functions in interval-censored multi-state models (Chapter 5 of Cook \& Lawless). \cite{Cook2020} For specific multi-state models, such as the (extended) illness-death model, estimators are available in the non-parametric \cite{Frydman1992, Frydman1995} as well as the flexible parametric setting. \cite{Joly2002} For general multi-state models, some parametric results are available in the semi-Markov framework. \cite{Aralis_2017, Aastveit_2023} For Markov models, parametric results are restricted to the (time-homogeneous) exponential distribution, allowing for (piecewise-)constant intensity functions \cite{Gentleman1994a} and are available in the \texttt{R} package \texttt{msm}. \cite{Jackson2011} This time-homogeneous approach is quite restrictive, yet numerical optimization is required to obtain estimates. For specific multi-state models, the computational load can be reduced as analytic expressions are available for the components of the likelihood function.  A flexible approach was proposed that allows for a mix of parametric and semi-parametric intensity functions using P-splines. \cite{Machado2018} This method, however, relies heavily on the time-homogeneous approach, as it approximates the chosen intensity functions by a fixed number of piecewise-constant intensity functions and therefore does not estimate the parameters directly. Titman \cite{Titman_2011} also proposed a flexible approach, modelling the intensities using B-splines. The estimates are then obtained by numerically solving the Kolmogorov forward equations, which can be computationally demanding in the presence of continuous covariates and experiences problems when the transition intensities have singularities. A non-parametric estimator for general interval-censored multi-state models was recently derived by Gu et al. \cite{Gu2023, Guthesis} using an EM algorithm with latent Poisson variables. The use of latent Poisson variables introduces some challenges, which we circumvent by taking a different approach.

In this article, we propose an Expectation Maximization algorithm to non-parametrically estimate the transition intensities in an interval-censored Markov multi-state model without loops. Without loops, the model does not allow for transitions to previously visited states.  We extend the algorithm to allow for transitions to an arbitrary but fixed number of states to be observed at exact times, facilitating the analysis of a mix of interval- and right-censored data. The notation, theory and EM algorithm are described in Section \ref{sec:methods}. In Section \ref{sec:EMalgorithm}, the EM algorithm is described in detail and a method to determine convergence of the algorithm is given. The latent Poisson estimator \cite{Gu2023} is described in Section \ref{sec:modelpoissondatalik} and used for comparison throughout the rest of the article. A simulation study is performed in Section \ref{sec:simulationstudy}, where the proposed estimator is compared with the latent Poisson \cite{Gu2023} and the time-homogeneous approach.\cite{Jackson2011} The three aforementioned methods are then used to analyse the Signal-Tandmobiel study \cite{Vanobbergen_2000} on tooth emergence in children. The article is concluded by a discussion and recommendations for future research.

\section{Methods}
\label{sec:methods}

In this section notation for multi-state models and data is introduced, as well as the proposed estimator. For further details on multi-state modelling, the reader is referred to Cook \& Lawless \cite{Cook2020} and Andersen \& Ravn.\cite{Andersen_Ravn_2024}

\subsection{Theory}
\label{sec:methodstheory}

Let $X(t)$ represent a time non-homogeneous Markov multi-state model with states in $\mathcal{H} = \{1, \ldots, H\}$ and denote the history (filtration) of the process up until time $t$ as $\mathcal{F}_{t} = \{X(u) ; 0 \leq u \leq t\}$. Not all transitions between states in $\mathcal{H}$ will be possible, we therefore let $\mathcal{V} = \{ (g,h) \in \mathcal{H} \times \mathcal{H}; \text{direct transition } g \to h  \text{ is possible} \}$ denote the set of possible direct transitions. We assume the model has no loops, implying the associated directed graph (see Figure 1) is acyclic. Without this assumption, a non-parametric estimator cannot be consistent as loops introduce uncertainty in the number of times a state is visited. Since we make no assumptions on the form of the intensities, we cannot compensate for this. For $(g, h) \in \mathcal{V}$ let the counting process $N_{gh}(t)$ denote the number of $g \to h$ transitions in $[0,t]$. As the counting processes for all transitions contain complete information on the process history, it can also be represented as $\mathcal{F}_{t} = \{N_{gh}(u) ; (g, h) \in \mathcal{V}, 0 \leq u \leq t \}$. 

The transition intensity, describing the instantaneous risk of a transition from state $g$ to $h$ with $(g, h) \in \mathcal{V}$ is given by:
\begin{align*}
    \alpha_{gh}(t) &= \lim_{dt \downarrow 0} \frac{\mathbb{P}(X(t + dt) = h \given X(t) = g, \mathcal{F}_{t-})}{dt} = \lim_{dt \downarrow 0} \frac{\mathbb{P}(X(t + dt) = h \given X(t) = g)}{dt},
\end{align*} where the second equality holds due to the Markov assumption. The cumulative transition intensity is then simply the integral $A_{gh}(t) = \int_0^t \alpha_{gh}(s) ds$. The process is only at risk of a transition at times when said transition can occur, therefore we also consider the transition intensity process:
\begin{align*}
    \lambda_{gh}(s) = \alpha_{gh}(s) Y_g(s),
\end{align*} with $Y_g(s) = \mathds{1}\{X(s-) = g\}$ the at-risk indicator for transitions out of state $g$ and $s-$ the time just before $s$. The intensity process is zero whenever the process is not at risk of a transition.  Let $C$ denote a right-censoring time after which the state of the process is no longer known. For interval-censored data, this is equal to the last time at which the state of the process is known. Additionally, we make the assumption of independent censoring:
\begin{align*}
    \alpha_{gh}(t) &=  \lim_{dt \downarrow 0} \frac{\mathbb{P}(X(t + dt) = h \given X(t) = g, C > t)}{dt} = \lim_{dt \downarrow 0} \frac{\mathbb{P}(X(t + dt) = h \given X(t) = g)}{dt}.
\end{align*} In other words, the knowledge that the process is currently uncensored does not influence the intensities.

Determining the probability of transitioning from one state to another is a vital component of multi-state modelling. Letting $0 \leq s \leq t$, this quantity is given by the transition probabilities:
\begin{align*}
    P_{gh}(s, t) = \mathbb{P}(X(t) = h \given X(s) = g).
\end{align*} We gather the cumulative intensities and transition probabilities into $H \times H$ matrices $\mathbf{A}(t) = \{ A_{gh}(t); g, h \in \mathcal{H}\}$ with diagonal entries $A_{gg}(t) = - \sum_{h \neq g} A_{gh}(t)$ and $\mathbf{P}(s, t) = \{P_{gh}(s, t); g, h \in \mathcal{H} \}$. For Markov multi-state models, the Chapman-Kolmogorov equations allow us to relate the transition probabilities to the transition intensities through product integration:
\begin{align} \label{eq:transprobprodint}
    \mathbf{P}(s,t) = \Prodi_{s < u \leq t} \left( \mathbf{I} + \textrm{d}\mathbf A(u) \right).
\end{align}

\subsection{Data}
\label{sec:methodsdata}

We consider interval-censored multi state data, where subjects $i = 1, \ldots, n$ are observed to be in state $x_{ij} \in \mathcal{H}$ at random (or possibly fixed) subject-specific visit times $T_{ij}$ with $j = 1, \ldots, n_i$. We denote $t_{ij}$ for the observed values of $T_{ij}$. Following Section $5.5$ of Cook \& Lawless \cite{Cook2020} we assume that the subject observation times follow a conditionally independent visit process:
\begin{align*}
    T_{ij} \perp \{X_i(s), s > t_{i,j-1}\} \given \mathcal{F}_{t_{i,j-1}, i},
\end{align*} with $\mathcal{F}_{t, i}$ the process history of individual $i$ up until time $t$. This means that the next visit time must not depend on the state of the process since the last visit time, but can depend on all information available at the last visit time.

\begin{figure}[!ht]
    \centering
    \includegraphics[width = 0.8 \textwidth]{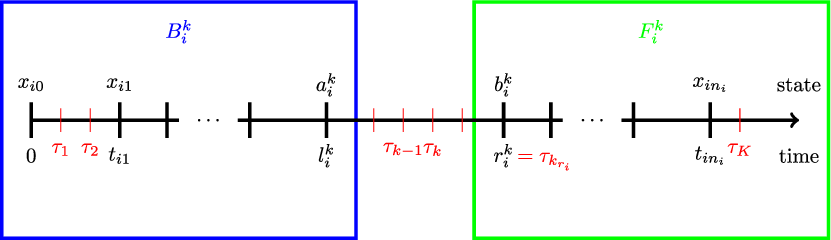}
    \caption{Graphical representation of data notation for a single subject $i$ and fixed $k \in \{1, \ldots, K\}$.}
    \label{fig:DataRepresentation_main}
\end{figure}

The observed set of information for subject $i$ is then given by:
\begin{align} \label{eq:observeddata}
    \mathbb{X}_i^\mathrm{O} = \{(x_{i0},t_{i0}),(x_{i1},t_{i1}),\ldots,(x_{in_i},t_{in_i})\}.
\end{align} For each subject we therefore observe realisations $X_i(t)$ of $X(t)$ such that $X_i(t_{ij}) = x_{ij}$. The observed data likelihood for interval-censored multi state data is then given by:
\begin{align} \label{eq:likpanelinterval}
    L^\mathrm{O} = \prod_{i=1}^n \prod_{j=1}^{n_i} P_{x_{i,j-1}x_{ij}}(t_{i,j-1}, t_{ij}).
\end{align} Let $0 = \tau_0 < \tau_1 < \ldots <\tau_K < \infty$ be the sorted unique time points of the set $\{t_{ij}$; $i=1,\ldots,n$, $j=1,\ldots,n_i$\}, and let $\mathcal{T} = \{ \tau_k; k = 1,\ldots,K \}$. We shall refer to an interval $(\tau_{k-1}, \tau_k]$ for an arbitrary $k$ as a \textbf{bin}. Assume that asymptotically the distance between unique observation times becomes arbitrarily small: $\lim_{n \to \infty}\max_k |\tau_{k}- \tau_{k-1}| = 0$. Define for each subject $i$ and state $g$ the at risk indicator $Y_{g,i}(s) = \mathds{1}\{ X_i(s-) = g \}$, indicating whether subject $i$ is in state $g$ just before time $s$ and let $N_{gh,i}(t)$ count the number of transitions $g \to h$ of subject $i$ in $[0,t]$. Note that the true state is only known at the observation times of the subjects. The value of $Y_{g,i}(s)$ is therefore only known right after the observation times. The number of transitions may be known if only a single path could have been taken by subject $i$, but this is not often the case. Similarly, we define $\lambda_{gh,i}(t) = \alpha_{gh}(t) Y_{g,i}(t)$ and let $dN_{gh,i}(t)$ be the observed jumps in $[t, t + dt)$ such that $\mathbb{P}(dN_{gh,i}(t) = 1 \given \mathcal{F}_{t-,i}) = d\Lambda_{gh,i}(t)$. For fixed $k \in \{1, \ldots, K\}$ we define:
\begin{align*}
    l_i^{k} = \max_{j = 1, \ldots, n_i} \{t_{ij}; t_{ij} \leq \tau_{k-1} \}, r_i^{k} = \min_{j = 1, \ldots, n_i} \{t_{ij}; t_{ij} \geq \tau_{k} \},
\end{align*} so that $(l_i^k, r_i^k]$ is the observational interval of subject $i$ containing $(\tau_{k-1}, \tau_k]$. Additionally, let $a_i^k = X_i(l_i^k)$ and $b_i^k = X_i(r_i^k)$ be the corresponding states occupied at the ends of this observational interval. We define $B_i^k \coloneqq \{ X_i(t_{ij}), t_{ij} \leq l_i^k \}$ and $F_i^k \coloneqq \{X_i(t_{ij}), t_{ij} \geq r_i^k\}$ as the past and future observations relative to the interval $[l_i^k, r_i^k]$. A graphical representation of this notation can be found in Figure \ref{fig:DataRepresentation_main}.

\subsection{EM algorithm}
\label{sec:methodsmodelintervalcens}

In this section we will non-parametrically estimate the cumulative transition intensities $A_{gh}(t)$ for interval-censored multi state data using the Expectation Maximisation (EM) algorithm.\cite{Dempster1977} The EM algorithm is an iterative procedure used to find maximum likelihood estimates in the presence of missing data. In the case of interval-censored data, the exact transition times between states are missing for individuals, warranting the use of this algorithm.

\subsubsection{Complete-data likelihood}
\label{sec:modelcompletedatalik}

To employ the EM algorithm, we must first determine the complete data likelihood. Suppose we observe all transitions for all subjects and therefore know the exact transition times $t^*_{ij}$ for $j = 1, \ldots, f_i$, with $f_i$ the number of transitions made by subject $i$. For each subject $i$ we then have the following set of observations:
\begin{align*}
    \mathbb{X}_i^\mathrm{C} = \{(s_{i0},t^*_{i0}),(s_{i1},t^*_{i1}),\ldots,(s_{if_i},t^*_{if_i})\},
\end{align*} with $X_i(t^*_{ij}) = s_{ij}$.
Under the assumption of independent censoring, the complete data likelihood contribution for a single subject is given by the canonical multinomial extension of the univariate likelihood (Section IV.4.1.5 of Andersen et al.):\cite{Andersen1997} 
\begin{align} \label{eq:liksinglecanmultinomial}
    L_i^\mathrm{C} &= \Prodi_{t \leq t^*_{if_i}} \prod_{g \in \mathcal{H}} \left\lbrace  \prod_{h \neq g} \left\lbrace \left( Y_{g,i}(t) dA_{gh}(t) \right)^{dN_{gh,i}(t)} \right\rbrace \left( 1 - \sum_{z \leftarrow g} dA_{gz}(t) \right)^{Y_{g,i}(t) - \sum_{z \leftarrow g}dN_{gz,i}(t)}  \right\rbrace,
\end{align} where $\sum_{z \leftarrow g}$ represents the sum over all states $z$ that can be reached directly (in a single transition) from $g$. This has the interpretation of conditionally independent multinomially distributed numbers of jumps from each state $g$, such that $(dN_{gh}(t): h \neq g) \sim \mathrm{Mult}(Y_{g,i}(t), dA_{gh}(t): h \neq g)$ given $\mathcal{F}_{t-,i}$.  For natural phenomena, it is reasonable to assume that the cumulative intensities are continuous. The product integral of the second term in Equation \eqref{eq:liksinglecanmultinomial} can then be approximated by an exponent, leading to the following approximation of the contribution to the likelihood:\cite{Cook2020, Andersen1997}
\begin{align}\label{eq:liksingle}
    L_i^\mathrm{C} = \left[ \Prodi_{t \leq t^*_{if_i}} \left\lbrace \prod_{(g,h) \in \mathcal{V}} \left( Y_{g,i}(t) dA_{gh}(t) \right)^{dN_{gh,i}(t)} \right\rbrace \right] \exp\left( - \sum_{(g,h) \in \mathcal{V}} \int_{0}^{t^*_{if_i}} Y_{g,i}(u) dA_{gh}(u) \right).
\end{align} This can then be interpreted as a ``Poisson'' approximation of the canonical multinomial likelihood such that $(dN_{gh,i}(t): h \neq g) \sim \mathrm{Pois}(Y_{g,i}(t) dA_{gh}(t): h \neq g)$ given $\mathcal{F}_{t-,i}$. Although it is possible to consider Equation \eqref{eq:liksinglecanmultinomial} for the EM-algorithm, the use of the ``Poisson'' approximation (Equation \eqref{eq:liksingle}) yields an estimator that is both easier to calculate and interpret. We therefore continue with Equation \eqref{eq:liksingle}, and with a slight abuse in naming refer to this likelihood as the ``multinomial'' complete data likelihood. The resulting estimator using the canonical multinomial extension (Equation \eqref{eq:liksinglecanmultinomial}) as well as for a ``multinoulli'' approximation can be found in Appendix \ref{sec:alternativeestimationprocedures}.

The log likelihood contribution for a single subject is then:
\begin{align} \label{eq:ell_i^c}
    \ell_i^\mathrm{C} \coloneqq \log(L_i^\mathrm{C}) &= \int_{0}^{t^*_{if_i}} \sum_{(g,h) \in \mathcal{V}} \Big[ \log\left(dA_{gh}(t) Y_{g,i}(t)\right) dN_{gh,i}(t) -  dA_{gh}(u) Y_{g,i}(u) du \Big].
\end{align}

In the complete-data setting, this expression will be maximized when the cumulative intensities are step functions with (possible) jumps at observed transition times. Although the exact transition times are not observed in the interval-censored setting, the NPMLE in the two-state survival setting has been shown to only have increments between some of the unique observation times.\cite{Turnbull1976} Later, a similar result was found for the progressive three state model \cite{Frydman1992} and the illness-death model.\cite{Frydman1995} The intervals where the cumulative intensities have increments are known as \emph{support intervals}. In general, the total value of the increments within these support intervals can be estimated, but the locations within the support intervals cannot. Turnbull \cite{Turnbull1976} therefore showed that the estimation problem could be reduced to the estimation of a right-continuous step function, with jumps at the right endpoints of the support intervals, with the aim of estimating the jump sizes at these endpoints. Since the estimation problem for general interval-censored multi-state models is more complex, we cannot expect to achieve better results than these methods. We therefore simplify the problem to estimating cumulative transition intensity increments at the right-endpoints of the bins, as the support intervals have been shown to depend completely on the unique observation times.

Following this line of reasoning, we restrict the estimation problem to cumulative intensity functions that can only make jumps at any of the unique observation times $\mathcal{T}$, similarly to Gu et al.\cite{Gu2023} This leads to an approximation of the complete-data log-likelihood, so that Equation \eqref{eq:ell_i^c} may be approximated by the sums:
\begin{align} \label{eq:ellic_approx}
    \ell_i^\mathrm{C} &= \sum_{k = 1}^K \sum_{(g,h) \in \mathcal{V}} \left\lbrace \log(\alpha_{gh}^k Y_{g,i}^k) d_{gh,i}^k -  \alpha_{gh}^k Y_{g,i}^k \right\rbrace = \sum_{k = 1}^K \sum_{(g,h) \in \mathcal{V}} \left\lbrace \log(\alpha_{gh}^k)  d_{gh,i}^k -  \alpha_{gh}^k Y_{g,i}^k \right\rbrace,
\end{align} where $\alpha_{gh}^k = d\overline{A}_{gh}(\tau_k)$, $Y_{g,i}^k = \overline{Y}_{g,i}(\tau_{k})$ and $d_{gh,i}^k = d\overline{N}_{gh,i}(\tau_k)$, with $\overline{A}_{gh}, \overline{Y}_{g,i}$ and $\overline{N}_{gh,i}$ the cumulative intensity, at-risk and counting processes in the simplified problem. These quantities then represent the jump of the cumulative intensity at $\tau_k$, presence in $g$ at $\tau_{k-1}$ and transitions made between $g$ and $h$ at $\tau_k$ respectively. The second equality in Equation \eqref{eq:ellic_approx} holds because $d_{gh,i}^k = 1$ implies $Y_{g,i}^k = 1$, and $d_{gh,i}^k = 1$ is only possible if $Y_{g,i}^k =1$. Now the approximate complete-data log likelihood is obtained by summing over all subjects:
\begin{align} \label{eq:loglikcomplete}
    \ell^\mathrm{C} = \sum_{i=1}^n \ell_i^\mathrm{C}  &= \sum_{k=1}^K \sum_{(g,h) \in \mathcal{V}} \left\lbrace d_{gh}^k \log(\alpha_{gh}^k) - \alpha_{gh}^k Y_{g}^k \right\rbrace,
\end{align} with $d_{gh}^k \coloneqq \sum_{i=1}^n d_{gh,i}^k$ and $Y_{g}^k \coloneqq \sum_{i=1}^n Y_{g,i}^k$. Although this is an approximation of the true complete-data likelihood given in Equation \eqref{eq:ell_i^c}, we will also refer to Equation \eqref{eq:loglikcomplete} as the complete-data likelihood throughout this article. It is important to keep in mind that this approximation is only reasonable if the assumption of diminishing distances between unique observation times holds.

Note that a support interval can consist of multiple bins and therefore the individual estimated jump sizes cannot be seen to represent the true increase in a single bin (although it is useful to think of them as such). Only the sum of the jumps over multiple bins that comprise a (to us unknown) support interval can be technically interpreted as an increment. 


\subsubsection{E-step}
\label{sec:modelcensestep}

In the E-step of the EM algorithm we calculate the expected value of the complete-data likelihood using a guess for the jumps in the cumulative intensities of all transitions. We then determine new estimates for these jump sizes in the M-step by maximising the complete data likelihood and use these estimates as guesses in the next iteration, repeating this until convergence. Denote by $\mathbf{\alpha} = (\alpha_{gh}^k; g \neq h \in \mathcal{H}, k = 1, \ldots, K)$ the vector of jumps in the cumulative intensities, and with $\widetilde{\mathbf{\alpha}} = (\widetilde{\alpha}_{gh}^k; g \neq h \in \mathcal{H}, k = 1, \ldots, K)$ the current estimate.  Let $\mathrm{O} = \{ \mathbb{X}_i^\mathrm{O}; i = 1, \ldots, n \}$ denote the observed data and $\mathrm{C} = \{\mathbb{X}_i^\mathrm{C}; i = 1, \ldots, n \}$ the complete data. 

The main goal is to calculate the conditional expectation of the complete data log-likelihood function:
\begin{align} 
    \mathbb{E}[ \ell^\mathrm{C} \given \mathrm{O}, \widetilde{\alpha}] &= \sum_{k=1}^K \sum_{(g,h) \in \mathcal{V}} \left\lbrace \mathbb{E}\left[d_{gh}^k  \given \mathrm{O}, \widetilde{\alpha} \right] \log(\alpha_{gh}^k) -  \alpha_{gh}^k \mathbb{E}\left[Y_{g}^k  \given \mathrm{O}, \widetilde{\alpha}  \right] \right\rbrace
\end{align} Note that this requires us to determine two conditional expectations.  It can be shown that:
\begin{align} \label{eq:paneldexpec}
    d_{gh}^k(\widetilde{\alpha}) \coloneqq \mathbb{E}\left[d_{gh}^k  \given \mathrm{O}, \widetilde{\alpha} \right] &=  \sum_{i=1}^n \frac{\widetilde{P}_{a_i^{k},g}(l_i^{k},\tau_{k-1}) \cdot \widetilde{\alpha}_{gh}^k \cdot \widetilde{P}_{h,b_i^{k}}(\tau_k,r_i^{k})}
    {\widetilde{P}_{a_i^{k},b_i^{k}}(l_i^{k},r_i^{k})}, \\
    Y_g^k(\widetilde{\alpha}) \coloneqq \mathbb{E}\left[Y_{g}^k  \given \mathrm{O}, \widetilde{\alpha}  \right] &=  \sum_{i=1}^n \frac{\widetilde{P}_{a_i^{k},g}(l_i^{k},\tau_{k-1}) \cdot \widetilde{P}_{g,b_i^{k}}(\tau_{k-1},r_i^{k})}{\widetilde{P}_{a_i^{k},b_i^{k}}(l_i^{k},r_i^{k})}, \label{eq:panelYexpec}
\end{align} with $\widetilde{\mathbf{P}}(s, t) = \prodi_{s < u \leq t} \left( \mathbf{I} + \textrm{d}\widetilde{\mathbf A}(u) \right)$ the product integral of the current estimates of the transition intensities. The derivation can be found in Appendix \ref{sec:EMderivationsPanel}.

\subsubsection{M-step}
\label{sec:modelcensmstep}

Having found the expected value of the likelihood function under the current estimates of the parameters, we need to maximise the resulting expected likelihood function. The jumps in the cumulative intensities are bounded, so the maximum must be found within a restricted optimisation space. Luckily, the complete data log-likelihood is a concave function and therefore we can use convex optimisation theory.\cite{Boyd_Vandenberghe_2023} 

The intensities $\alpha_{gh}^k$ represent conditional probabilities of making a transition in a single bin and therefore must be bounded by zero and one. Additionally, the total probability of leaving a state $g$ in any bin $k$ must also be smaller or equal than one. We therefore define the following optimisation region:
\begin{align*}
    C_\alpha &= \left\lbrace \alpha_{gh}^k, g \neq h \in \mathcal{H}, k = 1, \ldots, K; \alpha_{gh}^k \geq 0, \sum_{h \leftarrow g} \alpha_{gh}^k \leq 1 \right\rbrace,
\end{align*} Note that the two conditions $\alpha_{gh}^k \geq 0$ and $\sum_{h \leftarrow g} \alpha_{gh}^k \leq 1$ guarantee that $0 \leq \alpha_{gh}^k \leq 1$. We show in Appendix \ref{sec:EMderivationsPanel}  that the expected value of the complete data likelihood is maximised over the above region by the following expression:
\begin{align} \label{eq:alphaghksolution}  
    \alpha_{gh}^k &= \begin{cases}
        \frac{d_{gh}^k(\widetilde{\alpha})}{Y_g^k(\widetilde{\alpha})}, & \mu_g^k = 0, \\
        \frac{d_{gh}^k(\widetilde{\alpha})}{\sum_{h \leftarrow g} d_{gh}^k(\widetilde{\alpha})}, & \mu_g^k > 0,
    \end{cases}
\end{align} with 
\begin{align} \label{eq:mugk}
    \mu_g^k = \max\left(  0,  \sum_{h \leftarrow g}d_{gh}^k(\widetilde{\alpha}) - Y_g^k(\widetilde{\alpha}) \right).
\end{align} Having found this expression, we can employ the EM algorithm until a certain convergence criterion is met, see Algorithm \eqref{alg:NPMLE}.

\subsection{Including exactly observed transition times}
\label{sec:methodsmodelexactobs}

Until now we have considered the case of interval censored multi-state data. In multi-state models there are often states for which the entry time into the state is exactly observed (or right-censored). In the illness-death model for example, it is unlikely that the exact time of the occurrence of a disease is known but the exact time of death is almost always known. These ``exactly observed'' states are oftentimes absorbing, but we do not have to make this assumption. In this section we therefore consider the possibility of transitions into a fixed and known subset $\mathcal{E} \subseteq \mathcal{H}  $ to be observed at exact times, so that for an observation $(x_{ij}, t_{ij})$ with $x_{ij} \in \mathcal{E}$ this implies that the transition to $x_{ij}$ happened at $t_{ij}$. To derive the intensities with exactly observed states, we follow the same steps as in Section \ref{sec:methodsmodelintervalcens}.

Suppose we observe subject $i$ arriving in state $h \in \mathcal{E}$ at time $t_{ij}$ for some $j \in \{1, \ldots, n_i\}$. This means that the process must have been in a state $g \in \mathcal{H}$ which allows for a direct transition to $h$ right before time $t_{ij}$. Define $\mathcal{R}_{h} = \{g \in \mathcal{H}; \text{direct transition $g \to h$ is possible}\}$ as the states which $h \in \mathcal{H}$ is directly reachable from. As we assumed that the cumulative intensities have upward jumps only at times in $\mathcal{T}$, we therefore know that the subject must have been in $\mathcal{R}_h$ at time $t_{ij}-$. The observed data likelihood is now given by:
\begin{align} \label{eq:likpanelexact}
    L^{\mathrm{O}}_\mathrm{E} = \prod_{i=1}^n \prod_{j = 1}^{n_i} \left\lbrace P_{x_{i,j-1}x_{ij}}(t_{i,j-1}, t_{ij}) \right\rbrace^{1-\mathds{1}\{x_{ij} \in \mathcal{E}\}} \left\lbrace \sum_{m \in \mathcal{R}_{x_{ij}}} P_{x_{i,j-1}m}(t_{i,j-1}, t_{ij}-) dA_{mx_{ij}}(t_{ij}) \right\rbrace^{\mathds{1}\{x_{ij} \in \mathcal{E}\}},
\end{align} as opposed to Equation \eqref{eq:likpanelinterval}. Note that $t_{ij} = \tau_k$ for some $k \in \{1, \ldots, K\}$, therefore $t_{ij}- =  \tau_{k-1}$. On the other hand, the complete data still consists of all (exact) transition times, therefore the complete-data likelihood (Equation \eqref{eq:loglikcomplete}) does not change. In the E-step however, we now have additional information for the calculation of the expected values whenever a state in $\mathcal{E}$ is observed. Note that this additional information improves the variance of the estimation procedure, as there is more certainty about the states a subject can reside in before an exact transition and fewer possible transitions can have occurred at said time. Having many exactly observed states therefore always makes the estimation procedure ``easier'' and more precise.

\subsubsection{E-step}
\label{sec:modelexactestep}

To incorporate the newly available information, we re-evaluate the quantities $d_{gh}^k(\widetilde{\alpha})$ and $Y_{g}^k(\widetilde{\alpha})$. We show in Appendix \ref{sec:EMderivationsPanelExact} that when $b_i^k \in \mathcal{E}$ the conditional expectations are instead given by:
\begin{align} \label{eq:exactdexpec}
    d_{gh,i}^k(\widetilde{\alpha}) \coloneqq  \mathbb{E}\left[d_{gh,i}^k  \given \mathrm{O}, \widetilde{\alpha} \right] &= \frac{\widetilde{P}_{a_i^kg}(l_i^k, \tau_{k-1}) \cdot \widetilde{\alpha}_{gh}^k \cdot \sum_{m \in \mathcal{R}_{b_{i}^k}} \widetilde{\alpha}_{mb_i^k}^{k_{r_i}} \widetilde{P}_{hm}(\tau_k, \tau_{k_{r_i}-1}) }{\sum_{m \in \mathcal{R}_{b_{i}^k}} \widetilde{\alpha}_{mb_{i}^k}^{k_{r_i}} \widetilde{P}_{a_i^km}(l_i^k, \tau_{k_{r_i}-1})}, \\
    Y_{g,i}^k(\widetilde{\alpha}) \coloneqq \mathbb{E}\left[Y_{g,i}^k  \given \mathrm{O}, \widetilde{\alpha} \right] &= \frac{\widetilde{P}_{a_i^kg}(l_i^k, \tau_{k-1}) \cdot   \sum_{m \in \mathcal{R}_{b_{i}^k}} \widetilde{\alpha}_{mb_{i}^k}^{k_{r_i}} \widetilde{P}_{gm}(\tau_{k-1}, \tau_{k_{r_i}-1})  }{\sum_{m \in \mathcal{R}_{b_{i}^k}} \widetilde{\alpha}_{mb_{i}^k}^{k_{r_i}} \widetilde{P}_{a_i^km}(l_i^k, \tau_{k_{r_i}-1})}, \label{eq:exactYexpec}
\end{align} with $\tau_{k_{r_i}} \coloneqq r_i^{k}$, $d_{gh}^k(\widetilde{\alpha}) = \sum_{i=1}^n d_{gh,i}^k(\widetilde{\alpha})$ and $Y_{g}^k(\widetilde{\alpha}) = \sum_{i=1}^n Y_{g,i}^k(\widetilde{\alpha})$. When $b_i^k \notin \mathcal{E}$ the conditional expectations are still given by Equations \eqref{eq:paneldexpec} and \eqref{eq:panelYexpec}.

\subsubsection{M-step}
\label{sec:modelexactmstep}

For the M-step we would like to maximise the expected complete data likelihood function. Fortunately, nothing changes with respect to Section \ref{sec:modelcensmstep}, therefore we can still utilise Equation \eqref{eq:alphaghksolution} to update our estimate of the intensities.

\subsection{EM algorithm}
\label{sec:EMalgorithm}

The theory above suggests a way for the computation of the non-parametric maximum likelihood estimate of the intensities. An algorithm displaying the necessary steps to arrive at an estimate is given in Algorithm \ref{alg:NPMLE}.
\begin{algorithm}
\caption{Calculation of NPMLE of intensities in a MSM without loops.}\label{alg:NPMLE}
\textbf{Input: } Initial estimate $\widetilde{\alpha}$, Tolerance $\epsilon$. \\
\textbf{Output: } Final estimate $\alpha$, observed data log likelihood $\ell$. \\ 
\textbf{Initialize: } $\ell = 0$, $\ell_{\mathrm{old}} = -\infty$, $\alpha_{gh}^k = \infty$ if $(g, h) \in \mathcal{V}$ . 
\begin{algorithmic}
\Require $0 < \widetilde{\alpha}_{gh}^k \leq 1$ if $(g,h) \in \mathcal{V}$; $\epsilon > 0$
\While{$\max_{g,h,k} |\widetilde{\alpha}_{gh}^k - \alpha_{gh}^k| > \epsilon$}
\For{$g \in \mathcal{H}$}
\For{$h \neq g \in \mathcal{H}$}
\For{$k \in |\mathcal{T}|$}
\State Calculate $Y_{gh}^k(\widetilde{\alpha})$ using Equations \eqref{eq:panelYexpec} or \eqref{eq:exactYexpec}
\State Calculate $d_{gh}^k(\widetilde{\alpha})$ using Equations \eqref{eq:paneldexpec} or \eqref{eq:exactdexpec}
\State Calculate $\mu_g^k$ using Equation \eqref{eq:mugk}
\State Update $\widetilde{\alpha}_{gh}^k$ using Equation 
\eqref{eq:alphaghksolution}
\EndFor
\EndFor
\EndFor
\State $\ell_{\mathrm{old}} \gets \ell$
\State $\alpha \gets \widetilde{\alpha}$
\State Calculate $\ell$ using Equation \eqref{eq:likpanelinterval} or \eqref{eq:likpanelexact}
\EndWhile
\end{algorithmic}
\end{algorithm} \\
Algorithm \ref{alg:NPMLE} uses an intensity convergence criterion and therefore terminates when the largest change in any intensity is smaller than the tolerance. A different criterion could also be used, such as a likelihood convergence criterion: $|L - L_{\mathrm{old}}| > \epsilon$. Another approach can also be taken. For a convex optimisation problem, whenever an estimate $\widetilde{\alpha}$ satisfies the Karush-Kuhn-Tucker conditions (details in Appendix \ref{sec:EMderivationsPanel}) we are guaranteed to have arrived at a (local) maximum of the expected complete-data likelihood function (Section 5.5.3 of Boyd \& Vandenberghe).\cite{Boyd_Vandenberghe_2023} When this is the case, further iterations of the EM algorithm will not improve the objective function. This suggests a method to check whether the algorithm should be terminated, without keeping track of estimated quantities in previous iterations. Such a procedure is outlined in Algorithm \ref{alg:islocalmax}. The reduced gradient $\nabla_{gh}^k$ has the interpretation of how much the objective function can be improved by changing the corresponding estimate. Note that the reduced gradient can be calculated during every iteration of Algorithm \ref{alg:NPMLE} at little cost. Although Algorithm \ref{alg:islocalmax} suggests checking whether the reduced gradient is greater than $0$, it is also possible to stop when $\nabla_{gh}^k < \epsilon$ for all transitions and time bins.

\begin{algorithm}
\caption{Determine whether a (local) maximum has been reached.}\label{alg:islocalmax}
\textbf{Input: } Output of Algorithm \ref{alg:NPMLE}. \\
\textbf{Output: } TRUE/FALSE 
\begin{algorithmic}
\For{$g \in \mathcal{H}$}
\For{$h \neq g \in \mathcal{H}$}
\For{$k \in |\mathcal{T}|$}
\State Calculate $Y_{gh}^k(\widetilde{\alpha}), d_{gh}^k(\widetilde{\alpha}), \mu_g^k$ (see Algorithm \ref{alg:NPMLE})
\State Calculate the \textit{reduced gradient} $\nabla_{gh}^k = Y_g^k(\widetilde{\alpha}) - \frac{d_{gh}^k(\widetilde{\alpha})}{\widetilde{\alpha}_{gh}^k}  + \mu_g^k$
\EndFor
\EndFor
\EndFor
\If{$\nabla_{gh}^k > 0$ for any $g, h, k$}
\State return FALSE
\Else{ return TRUE}
\EndIf
\end{algorithmic}
\end{algorithm}

\subsection{Latent Poisson EM}
\label{sec:modelpoissondatalik}

Recently an EM approach for the NPMLE using latent Poisson variables has been developed,\cite{Gu2023} allowing for the inclusion of (time-dependent) covariates and random effects through a semi-parametric proportional hazards model. For comparison purposes, we consider their model without any covariates/random effects. 

The main difference between their approach and ours is the complete data likelihood employed to maximise the observed-data likelihood (Equation \eqref{eq:likpanelinterval}). In Section \ref{sec:modelcompletedatalik} we took the multinomial likelihood approach as described in Section 2.2 of Cook \& Lawless,\cite{Cook2020} whereas they take a latent Poisson approach. They also assume that the cumulative intensities only have jumps at the unique event times, but consider latent Poisson variables $W_{gh,i}^k$ for each bin instead of $d_{gh,i}^k$ as we do. These variables then indicate whether a certain transition in a bin could have happened (latently), and assign (non)-zero mass if so. They then show that the observed data likelihood using latent Poisson  variables is the same as Equation \eqref{eq:likpanelinterval}, therefore the observed-data likelihood can be maximised using the Poisson full data likelihood:
\begin{align} \label{eq:completedatalikpoisson}
    \ell^{\mathrm{C}}_{\mathrm{P}} &= \sum_{i=1}^n \left(  \sum_{k=1}^K \sum_{(g,h) \in \mathcal{V}} \mathds{1}\{ \tau_k \leq t_{in_i} \} \left[ W_{gh,i}^k \log(\alpha_{gh}^k) - \alpha_{gh}^k - \log(W_{gh,i}^k!)  \right]  \right)
\end{align} Performing an E- and M-step as we have, it is then possible to obtain an update rule for $\alpha$, which is described in more detail in Appendix \ref{sec:EMderivationslatentpoisson}.

Although possible in the latent Poisson setting, exactly observed event times are not considered for an arbitrary number of states \cite{Guthesis} referring to the inconsistency problems in mixed samples NPMLE estimation.\cite{Ma2012} Instead, they consider only the situation with a single absorbing state, for which observation times are known exactly and model the transition intensities into this state using B-splines. Due to the restriction to a single exactly observed state, we do not compare our model with the latent Poisson approach in the case of exactly observed states.


\section{Simulation study}
\label{sec:simulationstudy}

To investigate the performance of our proposed estimator we conduct a simulation study. Seeing as the latent Poisson EM estimator (Section \ref{sec:modelpoissondatalik}) has been shown to be consistent,\cite{Gu2023} we focus on comparing our method to this estimator. Quantities of primary interest are the bias and variance of the estimator with respect to the cumulative intensities and the transition probabilities as well as computation speed.

\subsection{Data generation}
\label{sec:simdatageneration}

We consider six scenarios in this simulation study. The aim of the scenarios is to investigate the effect of varying parameters of the data generation model. We first cover the parameters which are equal for all scenarios. We assume that all $n$ subjects are observed at the beginning of the study (time $0$) and that all observations are censored at the end of the study after $15$ years. We consider $n = 100, 300$ and $500$ subjects to assess the behaviour of the estimator with increasing samples. For each scenario and each number of subjects $n$ we create $N = 1000$ simulated data-sets.

The following simulation parameters are scenario specific. We consider the illness-death (ID) model, or the extended illness-death (EID) model (see Figure \ref{fig:MSmodelsgraphical}). Observed states are simulated assuming underlying Weibull($\lambda$, $k$) or Exponential($\lambda$) distributions for the transition times, with Weibull probability density function $f(x;\lambda, k) = \lambda k x^{k-1} e^{-\lambda x^k}$. Parameters were chosen so that the mean transition times were equal over all scenarios for comparable transitions. Subjects are either initially observed in the first (Alive) state or have an equal probability to start in the first or second (Illness) state. In some scenarios, death states are exactly observed. The inter-observation times (time between two consecutive observations for a subject) follow a uniform distribution with varying parameters.  Details can be found in Table \ref{table:simparams}.

\begin{table}[!ht]
\centering
\caption{Simulation parameters for the six considered scenarios.}
 \label{table:simparams}
 \resizebox{\textwidth}{!}{
 \begin{tabular}{@{} *{8}{c} @{}}
 & \multicolumn{4}{c}{ } & \multicolumn{3}{c}{Transition distributions}  \\
\cmidrule(l){6-8} Scenario & Model & Starting states & Exact states $\mathcal{E}$ & \makecell{(Inter-)observation \\ times} & $1 \to 2$ & $1 \to 3$ & $2 \to 3 \vee 2 \to 4$  \\
 \midrule
 1 & ID & 1 & $\emptyset$ & U$[0, 4.4]$ & Exp($0.1$) & Exp($0.05$) & Exp($0.1$) \\ 
 2 & ID & $\{1, 2\}$ & $\emptyset$ & U$[0, 4.4]$ & Exp($0.1$) & Exp($0.05$) & Exp($0.1$) \\
 3 & ID & 1 & $\emptyset$ & U$[0, 4.4]$ & Weib($0.5$, $\frac{1}{\sqrt{5}})$ & Weib$(0.5$, $\frac{1}{\sqrt{10}})$ & Weib$\left(2, \left(\frac{\Gamma(1.5)}{10}\right)^2 \right)$ \\ 
 4 & EID & 1 & $\{3, 4\}$ & U$[0, 4.4]$ &  Exp($0.1$) & Exp($0.05$) &  Exp($0.1$) \\
 5 & ID & $\{1, 2\}$ & $\emptyset$ & U$[2.9, 3.1]$ & Exp($0.1$) & Exp($0.05$) & Exp($0.1$) \\
 6 & ID & $\{1, 2\}$ & $\emptyset$ & U$[0, 2.44]$ \& U$[0, 7.33]$ & Exp($0.1$) & Exp($0.05$) & Exp($0.1$) \\

 \bottomrule
 \end{tabular}
 }
\end{table}

In the last two scenarios ($5$ and $6$) we investigate the effect of the observation schedule on the estimators. Scenario five represents the classical panel data schedule in chronological time. All subjects are scheduled to be observed once every $3$ years, with their response time deviating at most $0.1$ years in each direction uniformly.  In the sixth scenario, two different observation schemes are considered, with either a uniform$[0, 2.44]$ or uniform$[0, 7.33]$ distribution for the inter-observation times. This results in an average of $4$ or $10$ observations per subject in the $15$ year observation period, as opposed to the average of $6$ observations with uniform$[0, 4.4]$ inter-observation times.

For each of the scenarios and $n \in \{100, 300, 500\}$, we fit three multi-state models to each of the $N = 1000$ simulated data sets. The three considered models are the multinomial and latent Poisson EM approaches  as well as the time-homogeneous approach \cite{Jackson2011} using the \texttt{R} \cite{Rcore} package \texttt{msm}. The time-homogeneous approach assumes that transition intensities are constant over time per transition, which comes down to assuming exponential transition hazards. For scenarios $1, 2, 4$ and $5$ transition times are therefore generated under the time-homogeneous assumption. The initial estimates for the EM algorithms are chosen as $\widetilde{\alpha}_{gh}^k = \frac{1}{K}$ if $(g, h) \in \mathcal{V}$ and zero elsewhere and the algorithms are terminated when the change in all transition intensities is smaller than $0.001$. The time-homogeneous model is run with default parameters, letting the \texttt{msm} package determine initial estimates. 

To obtain performance measures we define the estimated cumulative intensities for the $v-$th data set as $\widehat{A}_{gh}^v(t)$ for all $(g, h) \in \mathcal{V}$ and transition probabilities $\widehat{P}_{gh}^v(t)$ for any tuple $(g,h)$. The true (oracle) values are denoted by $A_{gh}(t)$ and $P_{gh}(t)$. We then calculate the bias, variance and Root Mean Squared Error (RMSE) of the estimators over an equidistantly spaced partition of $[0, 15]$. For the cumulative intensities, fixing $(g,h) \in \mathcal{V}$ and $t \in \{0, 0.1, 0.2, \ldots, 15\}$ we calculate:
\begin{align}
    \mathrm{Bias}(t) &= \frac{\sum_{v=1}^{N} \left( \widehat{A}_{gh}^v(t) - A_{gh}(t) \right)}{N}, &  \mathrm{Var}(t) &= \frac{1}{N-1} \sum_{v=1}^N \left( \widehat{A}_{gh}^v(t) - \frac{\sum_{v=1}^N \widehat{A}_{gh}^v(t)}{N} \right)^2, 
\end{align} and $\mathrm{RMSE}(t) = \sqrt{\mathrm{Var}(t) + \mathrm{Bias}^2(t)}$. Transition probabilities are calculated by taking the product integral (Equation \eqref{eq:transprobprodint}) over the cumulative intensities.
\subsection{Results}
\label{sec:simresults}

For clarity we show only the results for $n = 500$, but the rest of the findings can be found in the Supplementary materials Section A.

\subsubsection{Scenario 1}
\label{sec:simscen1}

\begin{figure}[!ht]
    \centering
    \includegraphics[width = 0.9 \textwidth]{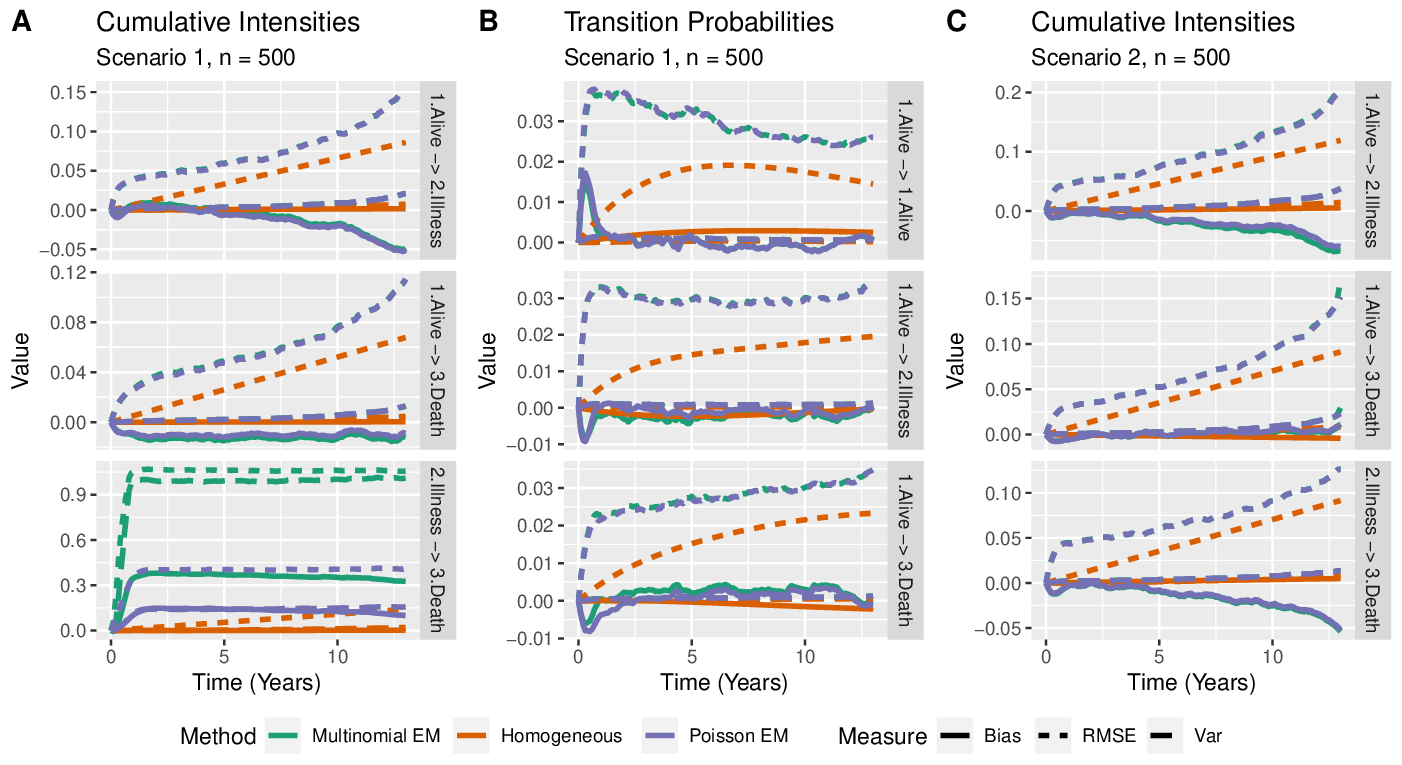}
    \caption{Bias, Variance and RMSE for $n = 500$ subjects of A) cumulative intensities in scenario 1, B) transition probabilities in scenario 1 and C) cumulative intensities in scenario 2.}
    \label{fig:sc1andsc2}
\end{figure}

In the first scenario, we consider the simple illness-death model with exponential transition hazards and all subjects starting in the alive state. Performance measures for the cumulative intensities of the three considered methods can be found in Figure \ref{fig:sc1andsc2} \textbf{A}. Overall, the time-homogeneous model performs best as it is correctly specified in this scenario. The multinomial and latent Poisson approach perform very similarly for transitions out of the alive state. For the transition from the illness state, the estimated cumulative intensity is very unstable in the first one to two years of the study for both EM approaches, but much more so for the multinomial approach. Ignoring the first two years however, the jumps in the cumulative intensities are recovered correctly as both the bias and variance curves remain relatively flat over time. As we are estimating conditional quantities (jumps in the cumulative intensities) instead of the marginal cumulative intensity function directly, we are interested in the slope of the performance curves rather than the value over time. The reason for the observed instability in the beginning of the study is the small number of subjects at risk of transitioning out of the illness state. The multinomial approach can then estimate a probability (jump in the cumulative intensity) close to one if only very few people are at risk and a subject makes a transition, explaining the jump of the variance curve to a value of around one. Similarly, as time progresses subjects will be absorbed in the death state, leaving less subjects for the estimation of intensities in all states. This explains the increase in the variability of the estimates over the study duration.

\begin{figure}[!ht]
    \centering
    \includegraphics[width = 0.9 \textwidth]{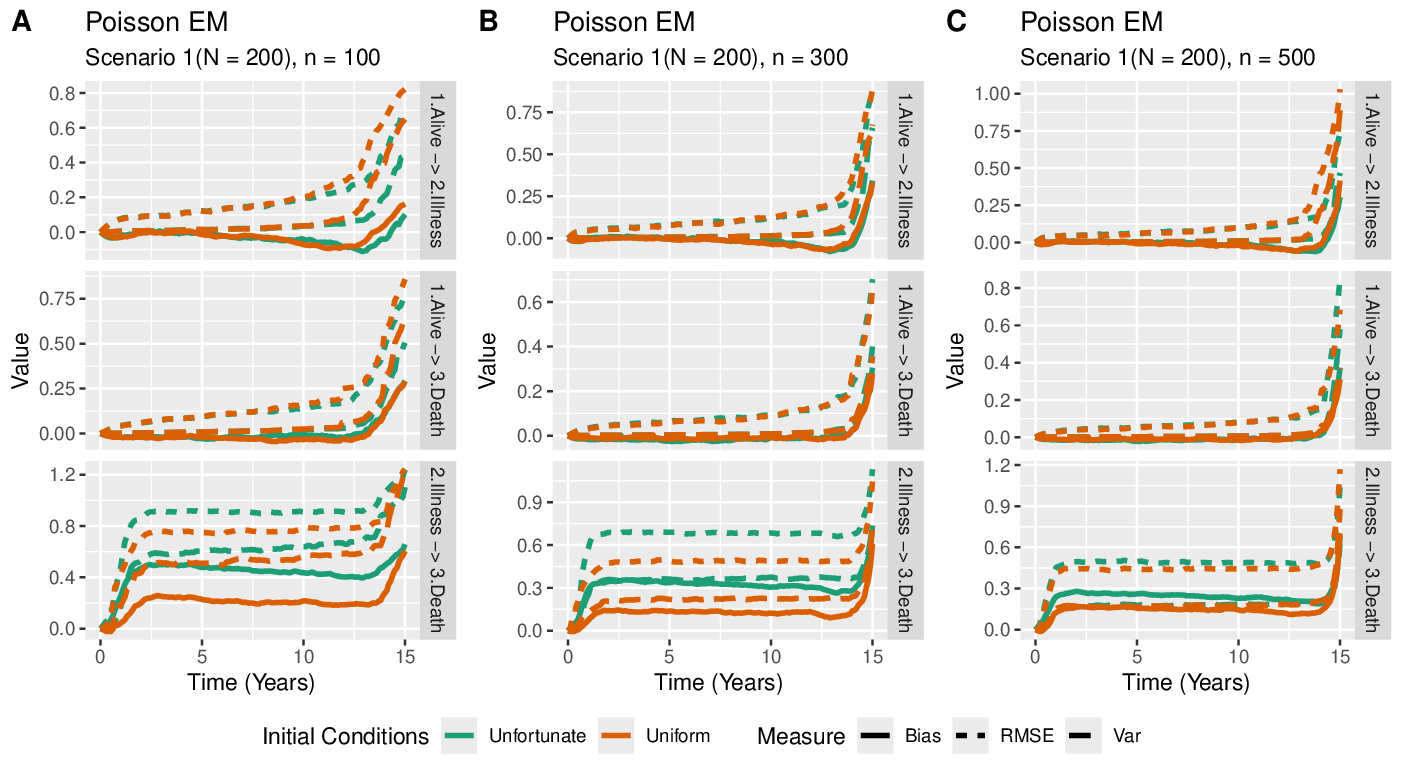}
    \caption{Bias, Variance and RMSE of cumulative intensities in scenario 1 with A) n = 100, B) n = 300, C) n = 500 subjects in $N = 200$ simulated data sets. Only Poisson EM is considered, with uniform and ``unfortunate'' initial conditions.}
    \label{fig:sc1wronginitintensities}
\end{figure}

The latent Poisson approach cannot change the estimates of its intensities for the $2 \to 3$ transition before a subject has entered state $2$ (see Appendix \ref{sec:poissoneminitialconditions}). Because of this, the initial bias for the Poisson approach in this scenario seems smaller, but is actually completely determined by the initial estimates. As the initial estimates were chosen uniformly depending on the number of unique observation times, the variance also results purely from the difference in observation times between data sets. To illustrate this occurrence, we fit the Poisson EM model on the first $200$ data-sets of this scenario using ``unfortunate'' initial estimates. These ``unfortunate'' initial estimates were chosen such that $\sum_{k=1}^K \alpha_{gh}^k = 1$ and $\sum_{k=1}^{\lfloor 0.1K \rfloor} \alpha_{gh}^k = 0.9$  for all $(g,h) \in \mathcal{V}$. In other words, we assigned $90$ percent of the initial ``mass'' to the first $10$ percent of the bins. A comparison of the results for uniform and ``unfortunate'' initial estimates can be found in Figure \ref{fig:sc1wronginitintensities}. Clearly, changing the initial estimates significantly changes the performance of the method for the $2 \to 3$ transition, which is a very undesirable property.

Oftentimes, transition probabilities are of greater interest than cumulative intensities. As all subjects start in the (first) alive state in this scenario, we take a look at the bias, variance and RMSE when recovering the transition probabilities from the first state $P_{1h}(t)$ for $h \in \{1, 2, 3\}$. Note that by considering the transition probabilities only from the first state, we can negate the problem of the small risk set encountered when considering the cumulative intensities. The results can be found in Figure \ref{fig:sc1andsc2} \textbf{B}. Unsurprisingly the time-homogeneous approach performs best. The multinomial and Poisson approach do not differ by much, but the bias for the probability of arriving in state $3$ (death) while starting in state $1$ (alive) is smaller in early time points for the multinomial approach, contrary to the observations for the cumulative intensities. Note that subjects can arrive in state $3$ through two different paths: either through illness ($1 \to 2 \to 3$) or directly ($1 \to 3$). Since transition probabilities are derived from estimated intensity jumps, the multinomial estimates, though seemingly more biased, outperform the latent Poisson estimates at early time points, which rely solely on initial guesses. All in all, both the Poisson and multinomial approach lead to biased estimates of the cumulative intensities for different reasons. Great care must be taken when interpreting estimates, as small risk sets can lead to unbalanced estimates for the multinomial approach. As the bias for the Poisson approach depends on the initial conditions, the multinomial estimator is preferred in models where subjects cannot start in every non-absorbing state.

\subsubsection{Scenario 2}
\label{sec:simscen2}

The Poisson EM estimate has been shown to be consistent for the cumulative intensities,\cite{Gu2023} but only when all non-absorbing states have a positive probability to be the initially observed state. In scenario $2$ both non-absorbing states are equally likely as starting states. We can therefore compare the performance of the multinomial estimator to a consistent estimator. Figure \ref{fig:sc1andsc2} \textbf{C} presents a comparison on the recovery of the underlying cumulative intensities.

There is not much difference between the Poisson and multinomial approach, and both perform well compared to the time-homogeneous approach. As the sample size $n$ increases, the RMSE of the EM estimators decreases at the same rate and they converge to almost identical estimates (see Supplementary materials Section A). It is therefore reasonable to assume that our proposed estimator is also consistent for the cumulative intensities when the non-absorbing states are covered in the initial observation.

\subsubsection{Scenario 3}
\label{sec:simscen3}

\begin{figure}[!ht]
    \centering
    \includegraphics[width = 0.9 \textwidth]{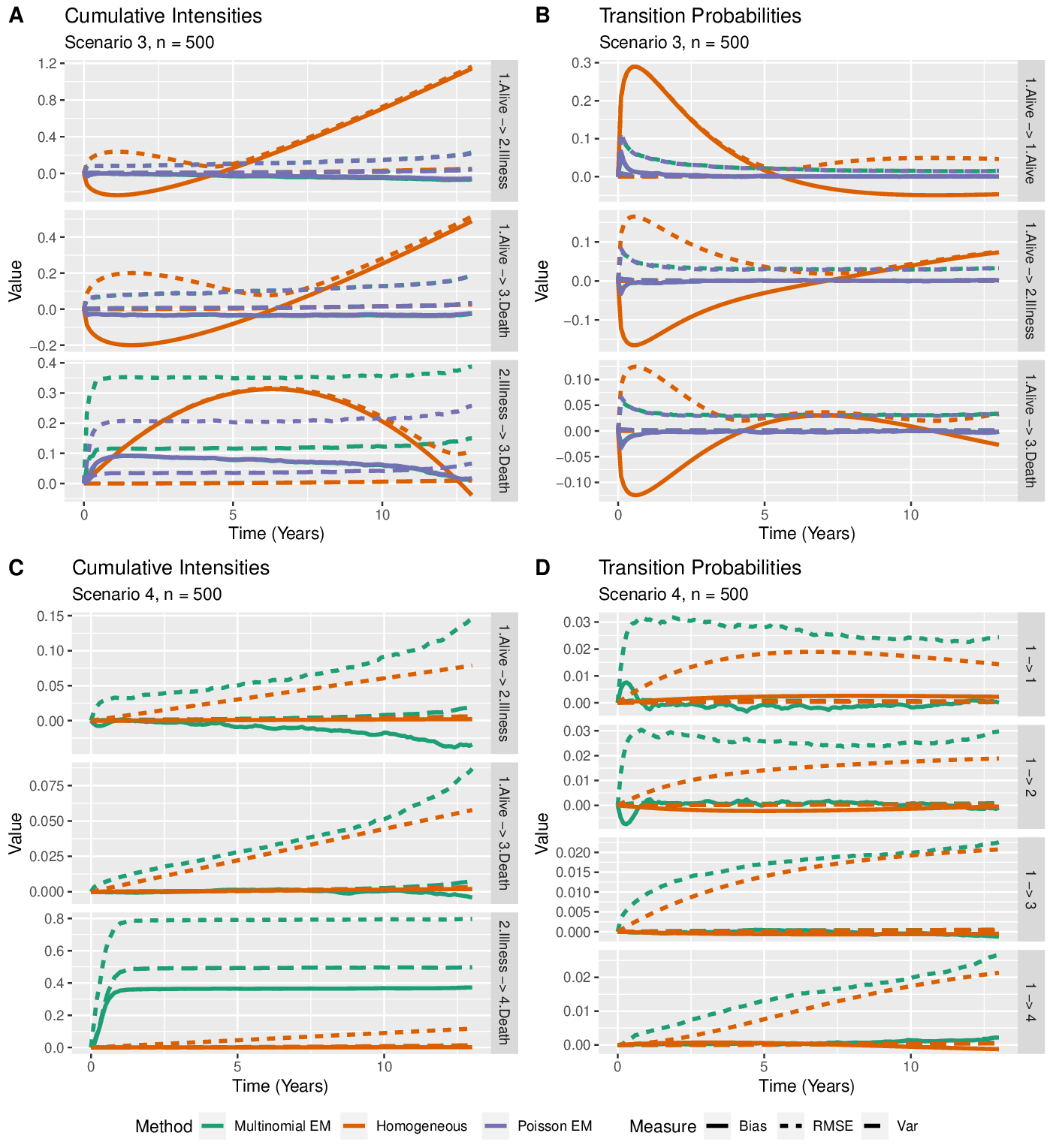}
    \caption{Bias, Variance and RMSE of cumulative intensities/transition probabilities for $n = 500$ subjects in A-B) scenario $3$ and C-D) scenario $4$.}
    \label{fig:sc3andsc4}
\end{figure}

In the previous scenarios, the data was generated under the time-homogeneous assumption. Now we take a look at how well the methods recover time varying intensities. The intensities for transitioning to illness and death were chosen to be decreasing functions of time, while the hazard from illness to death is increasing. In practice this means that subjects are more likely to become ill or pass away at the beginning of the study, and subjects that are ill for a long time are more likely to die. Performance measures for the cumulative intensities and transition probabilities for $n = 500$ can be found in Figure \ref{fig:sc3andsc4} \textbf{A-B}. 

As expected, the time-homogeneous model does not recover the underlying form of the hazards at all. We observe the same problem due to the small risk sets as in Scenario $1$, where the cumulative intensities seem to be more appropriately recovered by the latent Poisson approach, whereas the transition probabilities are recovered equally well by the two EM methods. All in all, the estimates are unreliable when only very few transitions are observed (early and late study times).

\subsubsection{Scenario 4}
\label{sec:simscen4}

In this scenario we consider the extended illness-death model with both death states exactly observed. We compare only the multinomial EM method with the time-homogeneous approach, as the latent Poisson approach was not worked out for this case. Note that the \texttt{msm} package allows for the inclusion of exactly observed states. The underlying hazards were therefore chosen to be exponential so that a comparison between the methods can be made. The results for $n=500$ can be found in Figure \ref{fig:sc3andsc4} \textbf{C-D}.

Both methods recover the transition probabilities from the initial state very well. The cumulative intensity estimate for the multinomial approach suffers from the low risk sets as before. The RMSE of the multinomial estimator is very similar to the well specified time-homogeneous model. In conclusion, we recover the intensities well even in the case of exactly observed states.

\subsubsection{Scenario 5 }
\label{sec:simscen56}

Scenario $5$ was considered to determine whether appropriate estimates could be recovered when classical panel data was considered in chronological time, resulting in no information being available on any subject for periods of at least $2.8$ years at a time. As a result, the distance between two consecutive unique observation times $|\tau_k - \tau_{k-1}|$ is very large for some $k$.  Figures summarizing the resulting performance of the estimators can be found in the Supplementary materials Section A. Clearly, the cumulative intensities are not recovered well, with the estimator oscillating around the true value over time. It is not possible to recover the true form of the intensities from such an estimator, which exhibits both positive and negative bias over time. The negative bias occurs due to the absence of any information, which is then overcompensated by a sudden influx of observations, leading to positive bias. With time-dependent intensities, this absence of information would make recovering the true form of the intensities even more difficult. The importance of the assumption that $\lim_{n \to \infty} \max_k |\tau_k - \tau_{k-1}| = 0$ becomes apparent, as it is violated in this scenario.

\subsubsection{Scenario 6}
\label{sec:simscen6}

The aim of scenario $6$ is to determine whether the variance of the estimators becomes smaller as the largest distance between adjacent unique observation times $\max_k |\tau_k - \tau_{k-1}|$ becomes smaller. This is achieved by considering two observation schemes where subjects are observed less and more frequently over time. The two different observation schemes are compared on a sample of $n = 500$ in the Supplementary materials Section $A$. Not surprisingly, the variance decreases as subjects are observed more frequently for all methods. On average, more frequent observations guarantee that the observation intervals for subjects will be smaller, making the estimation procedure ``easier''.

\subsubsection{Computation speed}
\label{sec:simcomputationspeed}

A critical attribute of a method is whether it can be used on real-life data sets. The time required to obtain estimates plays a major role therein. Both EM approaches require the calculation of the same number of product integrals per iterations, and these calculations make up the largest part of the total computation time. Each iteration therefore takes roughly the same time for both approaches. To compare the two approaches, the mean number of iterations (and standard deviation) required to reach convergence over the $N = 1000$ repeats in scenario $3$ is shown in Table \ref{table:iterations}. The Poisson approach requires significantly more iterations to reach convergence than our proposed method and was therefore much slower in our experience. To put time in perspective, it took the multinomial EM algorithm on average $6.5$ minutes to reach convergence on a data set with $n = 500$ using a single core of the $2.6$Ghz Intel Xeon Gold $6126$ processor as opposed to $21$ minutes for the Poisson algorithm.

\begin{table}[!ht]
\centering
 \caption{Mean number of iterations (standard deviation) required to reach convergence over the $N = 1000$ data sets in scenario $3$ for the two EM algorithms. }
 \label{table:iterations}
 \begin{tabular}{@{} *{4}{c} @{}}
 & \multicolumn{3}{c}{Sample size}  \\
\cmidrule(l){2-4} EM algorithm & $100$ & $300$ & $500$  \\
 \midrule
Multinomial & $331 (155)$ & $474 (162)$ & $513 (128)$ \\
Poisson & $549 (230)$ & $684 (202)$ & $702 (210)$ \\
 \bottomrule
 \end{tabular}
\end{table}

The time-homogeneous approach implemented in the \texttt{msm} package was extremely fast in the first three scenarios, taking less than a minute to run all simulations. In scenario $4$ however, the homogeneous approach took about twice as long as the multinomial EM algorithm. This happened because \texttt{msm} uses the standard optimisation routine in \texttt{R} to maximise the observed-data likelihood, which is very slow when the (derivative of the) likelihood becomes hard to evaluate.

\subsection{Key findings}

We summarize the findings from the simulation study:
\begin{itemize}
    \item The Poisson estimator only produces interpretable estimates whenever subjects can start in all non-absorbing states. If this is not the case, its estimates depend on the initial choice of intensities and the model should therefore be used with caution. 
    \item The multinomial approach always produces interpretable estimates, but special care must be taken as small risk sets can occur when not all non-absorbing states are covered, which will result in large variability in the estimates at the corresponding time points. As the number of subjects grows large, this problem is diminished. This problem does not affect transition probabilities out of states which are known to have a sufficient amount of subjects at risk. We therefore suggest to focus more on transition probabilities than cumulative intensities directly.
    \item When interpreting cumulative intensities, one must realise that the model does not directly estimate the marginal cumulative intensity, but the jumps in the cumulative intensities which are conditional quantities instead. 
    \item Computationally, the multinomial approach is faster than the latent Poisson approach and the time-homogeneous approach in the presence of exactly observed states. For purely interval-censored data with constant intensities, the time-homogeneous approach is best. Additionally, the time-homogeneous approach allows for the inclusion of covariates.
    \item The maximal distance between adjacent unique observation times must be sufficiently small for the EM approaches to yield interpretable estimates. As the observation schedule becomes more dense, the variance of the estimators shrinks
\end{itemize}
In practice this means that the time-homogeneous approach is preferred if the intensities are known to be constant and there are no exactly observed states. The multinomial EM approach should be considered for time-varying intensities and/or in the presence of exactly observed transitions. The Poisson EM approach should only be used if subjects can start in all non-absorbing states and a theoretical guarantee on consistency of the cumulative intensity is required. When there are large gaps in unique consecutive observation times, the non-parametric models will not recover the underlying intensity well and a model with a smoothness assumption is required.


\section{Application}
\label{sec:application}

In this section, we analyse a part of the Signal-Tandmobiel study (described in Vanobbergen et al.)\cite{Vanobbergen_2000} using an appropriate multi-state model. We are interested in investigating whether estimates obtained after fitting a truly complex multi-state model are reasonable.

\subsection{Signal Tooth study data description}
\label{sec:appdatadescription}

\begin{figure}[!ht]
\centering
    \includegraphics[width = 0.45\textwidth]{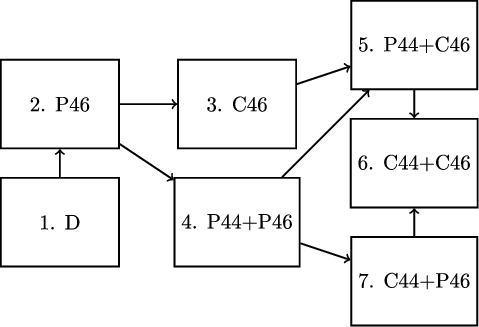}
    \caption{Multi-state models used for the analysis of the tooth data set. D stands for deciduous teeth, P for permanent and C for permanent tooth with Caries. The numbers indicate the respective location of the tooth. }
    \label{fig:MSMTooth}
\end{figure}

The Signal-Tandmobiel study contains longitudinal information on the occurrence of permanent teeth, caries as well as other (oral) descriptive characteristics between $1996$ and $2001$ for $4430$ children born in $1989$. The data set is freely available from the \texttt{R} package \texttt{icensBKL}.\cite{Bogaerts2020} Children were examined at most six times per year, with emergence and caries status being determined through inspection by a dentist leading to interval-censored data. All children start with only deciduous teeth (no permanent teeth). As the interest lies in a non-parametric approach, we leave out any covariate information. Instead, we focus on the emergence and caries status of spatially close permanent teeth $44$ and $46$. Clearly, a tooth must first emerge before caries can develop. We therefore consider the model shown in Figure \ref{fig:MSMTooth} B, with D indicating deciduous teeth only, P a permanent tooth and C a permanent tooth with caries. Only a single child had tooth $44$ emerge before tooth $46$, therefore this child was removed from the study in order to significantly simplify the model. Although no formal comparison can be made, we are interested in exploring the dynamics of caries occurrence and the emergence of two neighbouring permanent teeth. Additionally, we analyse the rate of caries occurrence on tooth $46$, comparing the pathways with and without the intermediary emergence of tooth $44$.

For the analysis, we fit the considered model using the multinomial and latent Poisson approach. The EM algorithms were run with initial estimates $\widetilde{\alpha}_{gh}^k = \frac{1}{K}$ if $(g, h) \in \mathcal{V}$ and zero elsewhere until the difference in the estimated intensities was smaller than $0.0001$. Additionally, a time-homogeneous model with piecewise-constant intensities was also considered, with initial estimates determined by the \texttt{msm} function. Due to numerical issues, only the time-homogeneous model was simplified to not include states $6$ and $7$. This simplification should not change the estimates by much, as less than $20$ children were observed in both states. As no transitions were observed before $6$ years of age, the cut-points for the homogeneous model were chosen at $6, 8$ and $10$ years to split the relevant time-frame into three equal parts.

\subsection{Results}
\label{sec:dataresults}

\begin{figure}[!ht]
    \centering
    \includegraphics[width = \textwidth]{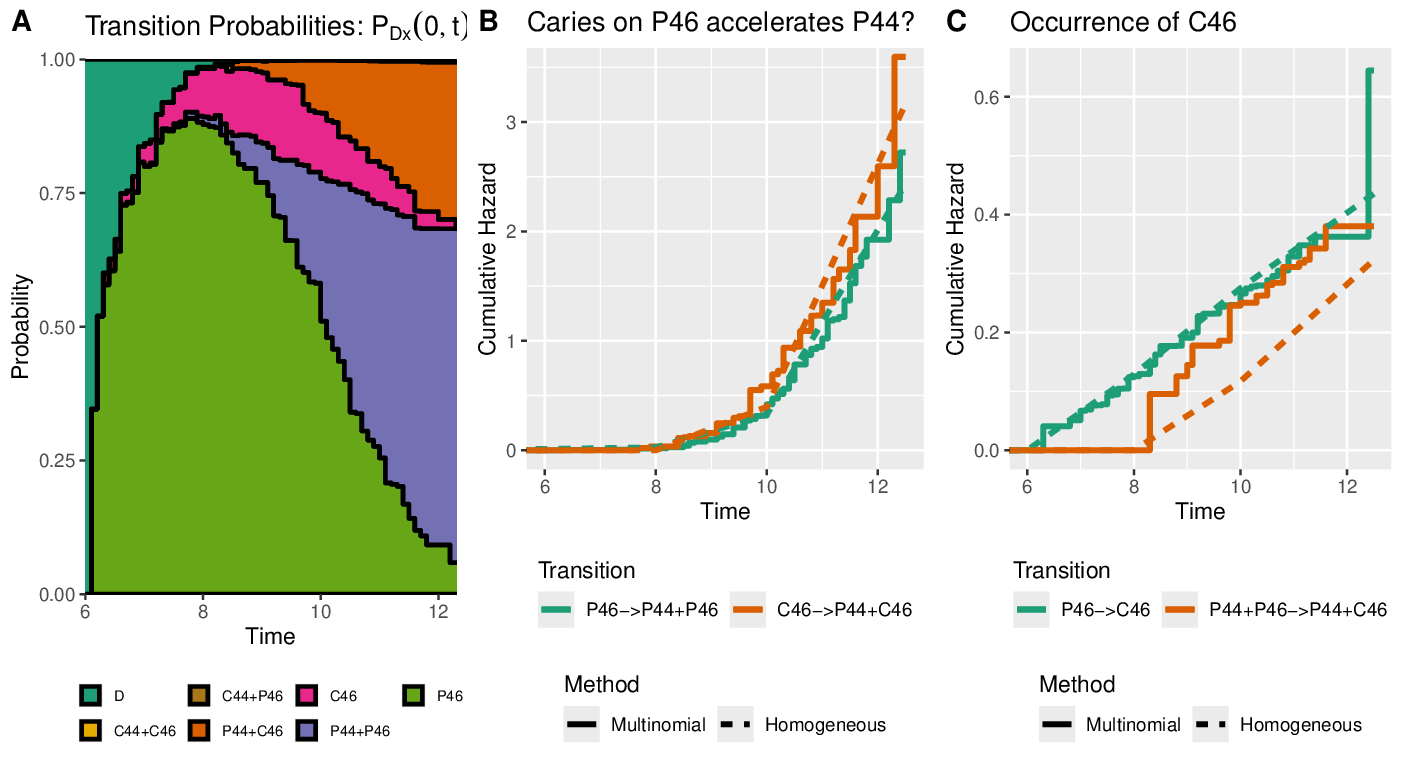}
    \caption{A) Transition probabilities (multinomial EM) from initial state D over time. The height of a colour indicates the probability of being in the corresponding state. B) Comparison of occurrence of permanent tooth $44$ with or without caries on permanent tooth $46$. C) Occurrence of caries on tooth $46$ through the two most common pathways. }
    \label{fig:cumhaztandmob}
\end{figure}

It took the multinomial algorithm $509$ iterations ($\approx 2$ hours) to converge, whereas the Poisson EM algorithm had not converged yet within the first $4000$ iterations and was stopped after $18$ hours. Although stopped prematurely, the resulting estimates did not differ much between the latent Poisson and multinomial approach. On further inspection, we found that the increase in the likelihood value per iteration was substantially smaller for the latent Poisson approach. Due to the small differences, we only show the results for the multinomial EM approach.  Transition probabilities from the initial state to all other states can be seen in Figure \ref{fig:cumhaztandmob} \textbf{A}. As mentioned before, tooth $46$ emerges before $44$, with caries on $46$ therefore often preceding the emergence of tooth $44$. Interestingly, almost none of the children developed caries on both teeth. Additionally, approximately $70$ percent of children did not develop caries on any of the $2$ considered teeth at the end of the study.

To investigate the dynamics of caries occurrence on tooth $46$ and the emergence of tooth $44$, we plot the cumulative hazards for the transitions P$46 \to $ P$44+$P$46$ and C$46 \to$ P$44+$C$46$ in Figure \ref{fig:cumhaztandmob} \textbf{B}. Note that C$46$ cannot occur before P$46$, and observing only C$46$ per definition implies that P$44$ has not happened yet therefore we expect the C$46 \to$ P$44+$C$46$ transition to have a delayed increase in the cumulative hazard. We can see however that the opposite is true for both models, with the cumulative hazard for the emergence of tooth $44$ being larger if caries is first observed on tooth $46$, giving rise to the belief that caries is associated with an accelerated occurrence of P$44$. This is perhaps a bit surprising, as an earlier analysis has shown that occlusal plaque (which usually precedes the occurrence of caries) on tooth $46$ has no effect on the emergence of tooth $44$ (see Example $4.2$ of Bogaerts et al.).\cite{Bogaerts2020}

The cumulative hazards of the two most common pathways to caries on tooth $46$ are shown in Figure \ref{fig:cumhaztandmob} \textbf{C}. A similar issue is also present here, as P$46$ must precede the caries-free state P$44+$P$46$ and therefore we expect the rate of caries occurrence from the second state to increase at a later time point. The overall slopes of the multinomial EM estimates suggest that neither of the two pathways considered is associated with an increased risk of developing caries. The observed increase in the slope for the second transition can be attributed to the smaller risk set at age 8, as a small risk set may lead to large jumps in the estimated cumulative hazard whenever caries is observed. The homogeneous estimates for the two transitions are comparable, with the two transitions displaying similar slopes but shifted in time. Although the same conclusion is reached using the two methods, the homogeneous approach requires the specification of cut-points. Such a choice requires manual analysis of the data set, in contrast to the EM approaches.

\section{Discussion}
\label{sec:discussion}

We have derived a non-parametric maximum likelihood estimator for the cumulative transition intensities in multi-state models without loops for a combination of interval- and right-censored data. The estimator makes use of the Expectation-maximisation algorithm and is based on the multinomial complete data likelihood for interval-censored multi state data. We present a necessary and sufficient condition to determine convergence of the algorithm to the true NPMLE. The estimated quantities allow to determine (cumulative) hazards for any transition in the model, as well as transition probabilities over the study period.

In comparison with the latent Poisson EM estimator,\cite{Gu2023} which was shown to be consistent for the cumulative intensities, our approach yielded almost identical estimates when the initial state can be any non-absorbing state. Additionally, our estimator yielded interpretable estimates for the transition probabilities when all subjects were initially observed in a single state, contrary to the latent Poisson approach. In such settings, small risk sets can lead to strong variability in the estimates. With larger sample sizes, this problem is mitigated. Finally, the multinomial EM algorithm was found to need significantly less iterations to converge to an estimate, making this approach computationally feasible.

Few approaches existed for the analysis of interval-censored multi-state data until recently, with the time-homogeneous model being the most popular. In contrast to the time-homogeneous approach, the non-parametric models allow the underlying cumulative intensities to be modelled with much greater flexibility. Additionally, in the presence of exactly observed (right-censored) states, our approach is computationally faster than existing time-homogeneous implementations as it does not rely on numerical optimisation techniques.

The multinomial and latent Poisson EM estimators allow for the proper analysis of interval-censored multi state data, negating the need to use inappropriate or incorrectly specified models. This in turn yields interpretable estimates and allows for unbiased conclusions in the presence of interval-censoring.

An important consideration when using the proposed estimator is the conditionally independent visit assumption. If future visit times are not independent of the state of the subject since the last visit, the proposed estimator might produce biased estimates. This problem may occur in the commonly used illness-death model, where sometimes subjects are more likely to visit a doctor shortly after contracting the disease.

The Markov assumption may not always hold. For exactly observed transition times, methods are available to test the Markov assumption \cite{Titman_Putter_2022} and to estimate state occupation and transition probabilities in the absence of the Markov assumption. \cite{Datta_Satten_2001, Putter_Spitoni_2016} For the interval-censored setting, no tests for the Markov assumption are available yet and alternative methods are lacking as well. Testing the Markov assumption is much more complex in this setting due to the lack of information about actual transition times.

A condition was derived to determine whether the algorithm has converged towards a (local) maximum. An important related problem is determining whether the obtained estimate is the NPMLE. Gentleman \& Geyer \cite{Gentleman1994} have provided conditions to check this for the interval-censored survival setting. Their result can be extended to the multi-state setting by considering the KKT conditions (see Section \ref{sec:EMderivationsPanel}) for the observed-data likelihood function instead of the expected complete-data likelihood function (Algorithm \ref{alg:islocalmax}). This would require the calculation of the derivative of the observed-likelihood function as well as an expression for the Lagrange multipliers (See Proposition \ref{prop:KKTconditions}). Although the derivative may be approximated numerically, obtaining an expression for the Lagrange multipliers may be very challenging. Another issue that can arise is that the NPMLE might not exist or not be unique.\cite{Gentleman1994, Hudgens2001} The NPMLE may not be unique when the observed data log-likelihood function is not strictly concave. For the standard survival setting, a sufficient condition for the uniqueness of the NPMLE is given in Gentleman \& Geyer.\cite{Gentleman1994} A similar condition could be derived for interval-censored multi-state data as well.

The proposed estimator yields virtually identical estimates as the latent Poisson approach, thereby implying consistent estimation of the cumulative intensities. A theoretical confirmation of this presumption could be derived following the procedure in Gu et al. \cite{Gu2023} Furthermore, the potential extension of our estimator to the semi-parametric framework via a proportional hazards approach should be considered. In a semi-parametric approach, a hypothesis test of a zero regression coefficient has been used to compare survival curves in an interval-censored setting.\cite{Finkelstein1986} A similar approach could also be used for interval-censored multi state data, and would allow us to formally compare the cumulative intensities in the Signal Tandmobiel study. 

Asymptotic properties of the introduced estimator were not discussed in this article. We discuss a few results for the simple two-state survival model with interval-censored data (see Chapter 3 of Sun). \cite{Sun_2011} First of all, the NPMLE of the distribution function is known to have only $n^{1/3}$-convergence rate and does not have an asymptotic normal distribution. When a certain proportion of failure times is exactly observed (plus some regularity conditions hold) or the support of the distribution function is discrete, the usual $n^{1/2}$-convergence rate holds, and the resulting asymptotic distribution is normal. Although possible to construct pointwise confidence limits in all of the above situations, either estimation of non-trivial quantities or the asymptotic variance of the distribution is required, which has no closed form. Due to these difficulties, confidence limits in the two-state setting are usually determined using bootstrap methods due to the low computational cost. Similar results are likely to hold in the multi-state setting. Hudgens et al. \cite{Hudgens2001} made the same presumption for the interval-censored competing risks model: ``If the support of the censoring mechanism is discrete and finite, the estimation of the cumulative incidence functions becomes a finite dimensional estimation problem and we expect the NPMLE to have the usual $n^{1/2}$ convergence rate. If the random variables dictating the censoring are treated as continuous, the rate of convergence of the NPMLE will likely not be $n^{1/2}$, and derivation of the limiting distribution will not be trivial.'' Part of these presumptions were later reinforced by Gu (2023), \cite{Guthesis}  who has shown the latent Poisson estimator to have a convergence rate of $n^{1/3}$ in the case of continuously distributed censoring. This suggests the use of bootstrap methods to obtain confidence limits, which is unfortunately not feasible due to the computational complexity of the models.

In interval-censored data, the state of a subject at a visit time is sometimes only known to lie in a set of possible states. It is possible to consider this extension to our model as well, following existing results for the time-homogeneous model.\cite{Jackson2011}  This would allow us to consider a different right-censoring mechanism, where the right-censoring time could also be a time at which the state of a subject is known to not lie in any of the exactly observed states. Furthermore, extensions for truncated data should be considered.

\section{Acknowledgements}

This work was performed using the ALICE compute resources provided by Leiden University.

\section{Software}

The methods described in this article are implemented in the \texttt{R} package \texttt{icmstate}.\cite{icmstateR} The code used to perform the data analysis and simulation study is available on GitHub. \cite{githubicmstate}

\bibliographystyle{unsrt}  
\bibliography{references}

\begin{thebibliography}{10}

\bibitem{Bogaerts2020}
K.~Bogaerts, A.~Komárek, and E.~Lesaffre.
\newblock {\em Survival Analysis with Interval-Censored data}.
\newblock Taylor \& Francis Group, Boca Raton, FL, 2020.

\bibitem{Turnbull1976}
B.~W. Turnbull.
\newblock The empirical distribution function with arbitrarily grouped, censored and truncated data.
\newblock {\em J R Stat Soc Series B Stat Methodol}, 38(3):290--295, July 1976.

\bibitem{Lindsey1998}
J.~C. Lindsey and L.~M. Ryan.
\newblock Methods for interval-censored data.
\newblock {\em Stat Med}, 17(2):219--238, January 1998.

\bibitem{Commenges2002}
D.~Commenges.
\newblock Inference for multi-state models from interval-censored data.
\newblock {\em Stat Methods Med Res}, 11(2):167--182, April 2002.

\bibitem{AalenJohansen1978}
O.~O. Aalen and S.~Johansen.
\newblock An empirical transition matrix for non-homogeneous {M}arkov chains based on censored observations.
\newblock {\em Scand J Stat}, 5(3):141--150, 1978.

\bibitem{Cook2020}
R.~J. Cook and J.~F. Lawless.
\newblock {\em Multistate Models for the Analysis of Life History Data}.
\newblock Taylor \& Francis Group, Boca Raton, FL, 2020.

\bibitem{Andersen_Ravn_2024}
P.~K. Andersen and H.~Ravn.
\newblock {\em Models for Multi-State Survival Data: Rates, Risks, and Pseudo-Values}.
\newblock Taylor \& Francis Group, Boca Raton, FL, 2024.

\bibitem{Putter2006}
H.~Putter, M.~Fiocco, and R.~B. Geskus.
\newblock Tutorial in biostatistics: competing risks and multi‐state models.
\newblock {\em Stat Med}, 26(11):2389--2430, October 2006.

\bibitem{SHARE}
SHARE.
\newblock Survey of health, ageing and retirement in europe.
\newblock {\url{https://share-eric.eu/}}.
\newblock Accessed: March 4, 2025.

\bibitem{AHEAD}
HRS.
\newblock University of michigan health and retirement study.
\newblock {\url{https://hrs.isr.umich.edu/}}.
\newblock Accessed: March 4, 2025.

\bibitem{Frydman1992}
H.~Frydman.
\newblock A nonparametric estimation procedure for a periodically observed three‐state {M}arkov process, with application to {AIDS}.
\newblock {\em J R Stat Soc Series B Stat Methodol}, 54(3):853--866, July 1992.

\bibitem{Frydman1995}
H.~Frydman.
\newblock Nonparametric estimation of a {M}arkov `illness-death’ process from interval-censored observations, with application to diabetes survival data.
\newblock {\em Biometrika}, 82(4):773--789, December 1995.

\bibitem{Joly2002}
P.~Joly.
\newblock A penalized likelihood approach for an illness-death model with interval-censored data: application to age-specific incidence of dementia.
\newblock {\em Biostatistics}, 3(3):433--443, September 2002.

\bibitem{Aralis_2017}
Hilary Aralis and Ron Brookmeyer.
\newblock A stochastic estimation procedure for intermittently-observed semi-markov multistate models with back transitions.
\newblock {\em Stat Methods Med Res}, 28(3):770--787, Nov 2017.

\bibitem{Aastveit_2023}
Marthe~Elisabeth Aastveit, C\'{e}line Cunen, and Nils~Lid Hjort.
\newblock A new framework for semi-{M}arkovian parametric multi-state models with interval censoring.
\newblock {\em Stat Methods Med Res}, 32(6):1100--1123, Apr 2023.

\bibitem{Gentleman1994a}
R.~C. Gentleman, J.~F. Lawless, J.~C. Lindsey, and P.~Yan.
\newblock Multi‐state {M}arkov models for analysing incomplete disease history data with illustrations for {HIV} disease.
\newblock {\em Stat Med}, 13(8):805--821, April 1994.

\bibitem{Jackson2011}
C.~H. Jackson.
\newblock Multi-state models for panel data: The msm package for {R}.
\newblock {\em J Stat Softw}, 38(8):1--28, 2011.

\bibitem{Machado2018}
R.~J.~M. Machado and A.~{van den Hout}.
\newblock Flexible multistate models for interval-censored data: Specification, estimation, and an application to ageing research.
\newblock {\em Stat Med}, 37(10):1636--1649, January 2018.

\bibitem{Titman_2011}
Andrew~C. Titman.
\newblock Flexible nonhomogeneous markov models for panel observed data.
\newblock {\em Biometrics}, 67(3):780--787, Feb 2011.

\bibitem{Gu2023}
Y.~Gu, D.~Zeng, G.~Heiss, and D.~Y. Lin.
\newblock Maximum likelihood estimation for semiparametric regression models with interval-censored multistate data.
\newblock {\em Biometrika}, 111(3):971--988, November 2024.

\bibitem{Guthesis}
Y.~Gu.
\newblock {\em Statistical methods for analyzing interval-censored multi-state data}.
\newblock PhD thesis, University of North Carolina at Chapel Hill, Chapel Hill, North Carolina, US, 2023.

\bibitem{Vanobbergen_2000}
J.~Vanobbergen, L.~Martens, E.~Lesaffre, and D.~Declerck.
\newblock The {Signal-Tandmobiel}\textregistered \text{ } project, a longitudinal intervention health promotion study in {Flanders (Belgium)}: baseline and first year results.
\newblock {\em Eur J Paediatr Dent}, 2:87--96, 2000.

\bibitem{Dempster1977}
A.~P. Dempster, N.~M. Laird, and D.~B. Rubin.
\newblock Maximum likelihood from incomplete data via the {EM} algorithm.
\newblock {\em J R Stat Soc Series B Stat Methodol}, 39(1):1--22, September 1977.

\bibitem{Andersen1997}
P.~K. Andersen, {\O}rnulf Borgan, R.~D. Gill, and N.~Keiding.
\newblock {\em {Statistical Models Based on Counting Processes}}.
\newblock Springer, New York, NY, 1997.

\bibitem{Boyd_Vandenberghe_2023}
S.~P. Boyd and L.~Vandenberghe.
\newblock {\em Convex optimization}.
\newblock Cambridge University Press, 2023.

\bibitem{Ma2012}
Y.~Ma and Y.~Wang.
\newblock Efficient distribution estimation for data with unobserved sub-population identifiers.
\newblock {\em Electron J Stat}, 6:710--737, Jan 2012.

\bibitem{Rcore}
{R Core Team}.
\newblock {\em R: A Language and Environment for Statistical Computing}.
\newblock R Foundation for Statistical Computing, version 4.2.0, 2025.

\bibitem{Titman_Putter_2022}
Andrew~C Titman and Hein Putter.
\newblock General tests of the {M}arkov property in multi-state models.
\newblock {\em Biostatistics}, 23(2):380--396, Apr 2022.

\bibitem{Datta_Satten_2001}
Somnath Datta and Glen~A. Satten.
\newblock Validity of the {Aalen–Johansen} estimators of stage occupation probabilities and {Nelson–Aalen} estimators of integrated transition hazards for non-markov models.
\newblock {\em Stat Probab Lett}, 55(4):403--411, Dec 2001.

\bibitem{Putter_Spitoni_2016}
Hein Putter and Cristian Spitoni.
\newblock Non-parametric estimation of transition probabilities in non-{M}arkov multi-state models: The landmark aalen–johansen estimator.
\newblock {\em Stat Methods Med Res}, 27(7):2081--2092, Oct 2016.

\bibitem{Gentleman1994}
R~Gentleman and C.~J. Geyer.
\newblock Maximum likelihood for interval censored data: Consistency and computation.
\newblock {\em Biometrika}, 81(3):618--623, 1994.

\bibitem{Hudgens2001}
M.~G. Hudgens, G.~A. Satten, and I.~M. Longini.
\newblock Nonparametric maximum likelihood estimation for competing risks survival data subject to interval censoring and truncation.
\newblock {\em Biometrics}, 57(1):74--80, March 2001.

\bibitem{Finkelstein1986}
D.~M. Finkelstein.
\newblock A proportional hazards model for interval-censored failure time data.
\newblock {\em Biometrics}, 42(4):845--854, December 1986.

\bibitem{Sun_2011}
Jianguo Sun.
\newblock {\em The Statistical Analysis of Interval-Censored Failure Time Data}.
\newblock Springer, New York, NY, 2011.

\bibitem{icmstateR}
Daniel Gomon and Hein Putter.
\newblock icmstate: Interval censored multi-state models.
\newblock \url{https://cran.r-project.org/package=icmstate}, 2024.
\newblock R package version 0.1.1.

\bibitem{githubicmstate}
Daniel Gomon and Hein Putter.
\newblock icmstate.
\newblock \url{https://github.com/d-gomon/icmstate}, 2024.
\newblock Accessed: March 4, 2025.

\end{thebibliography}

\appendix

\section{Derivations for the EM algorithm}
\label{sec:EMderivations}

\subsection{Interval-censored multi state data}
\label{sec:EMderivationsPanel}

\begin{proposition} \label{prop:condexpecinterval}
Consider the conditional expectation of the complete data log-likelihood function for interval-censored multi state data:
\begin{align} \label{eq:condexpecll}
    \mathbb{E}[ \ell^\mathrm{C} \given \mathrm{O}, \widetilde{\alpha}] &= \sum_{k=1}^K \sum_{(g,h) \in \mathcal{V}} \left\lbrace \mathbb{E}\left[d_{gh}^k  \given \mathrm{O}, \widetilde{\alpha} \right] \log(\alpha_{gh}^k) -  \alpha_{gh}^k \mathbb{E}\left[Y_{g}^k  \given \mathrm{O}, \widetilde{\alpha}  \right] \right\rbrace.
\end{align} The following conditional expectations when no transition times are known exactly are given by:
\begin{align*}
    d_{gh}^k(\widetilde{\alpha}) \coloneqq \mathbb{E}\left[d_{gh}^k  \given \mathrm{O}, \widetilde{\alpha} \right] &=  \sum_{i=1}^n \frac{\widetilde{P}_{a_i^{k},g}(l_i^{k},\tau_{k-1}) \cdot \widetilde{\alpha}_{gh}^k \cdot \widetilde{P}_{h,b_i^{k}}(\tau_k,r_i^{k})}
    {\widetilde{P}_{a_i^{k},b_i^{k}}(l_i^{k},r_i^{k})}, \\
    Y_g^k(\widetilde{\alpha}) \coloneqq \mathbb{E}\left[Y_{g}^k  \given \mathrm{O}, \widetilde{\alpha}  \right] &=  \sum_{i=1}^n \frac{\widetilde{P}_{a_i^{k},g}(l_i^{k},\tau_{k-1}) \cdot \widetilde{P}_{g,b_i^{k}(\tau_{k}-,r_i^{k})}}{\widetilde{P}_{a_i^{k},b_i^{k}}(l_i^{k},r_i^{k})},
\end{align*} with $\widetilde{\mathbf{P}}(s, t) = \prodi_{s < u \leq t} \left( \mathbf{I} + \textrm{d}\widetilde{\mathbf{A}}(u) \right)$ the product integral of the current estimates of the cumulative transition intensities.
\end{proposition}

\begin{proof}
For ease of notation, we define $\widetilde{\mathbb{E}}[\cdot] = \mathbb{E}[\cdot \given  \widetilde{\alpha}]$ and $\widetilde{\mathbb{P}}[\cdot] = \mathbb{P}[\cdot \given  \widetilde{\alpha}]$ as the conditional expectation and probability given the current estimates of the cumulative transition intensities. As we have assumed that the cumulative intensity is a right-continuous step function with jumps only in $\mathcal{T}$, we have:
\begin{align*}
    \mathbb{E}[d_{gh}^k  \given O, \widetilde{\alpha} ] &= \sum_{i=1}^n \mathbb{E}[\mathds{1}\{X_i(\tau_k) = h , X_i(\tau_k-) = g\} \given O, \widetilde{\alpha}] \\
    &= \sum_{i=1}^n \mathbb{P}(X_i(\tau_k) = h, X_i(\tau_k-) = g \given \mathbb{X}_i^\mathrm{O}, \widetilde{\alpha}) \eqqcolon \sum_{i=1}^n d_{gh,i}^k(\widetilde{\alpha}) \eqqcolon d_{gh}^k(\widetilde{\alpha}), \\
    \mathbb{E}\left[Y_{g}^k  \given O, \widetilde{\alpha} \right] &= \sum_{i=1}^n \mathbb{E}[\mathds{1}\{X_i(\tau_k-) = g \} \given  O, \widetilde{\alpha}] \\
    &= \sum_{i=1}^n \mathbb{P}(X_i(\tau_k-) = g \given \mathbb{X}_i^\mathrm{O}, \widetilde{\alpha}) \eqqcolon \sum_{n=1}^n Y_{g,i}^k(\widetilde{\alpha})  \eqqcolon Y_g^k(\widetilde{\alpha}), 
\end{align*} with $\tau_k-$ the time just before $\tau_k$ and $\mathbb{X}_i^\mathrm{O}$ the observed information for subject $i$ as in Section \ref{sec:methodsdata}. Fix some $k \in \{1, \ldots, K\}$ and $i \in \{1, \ldots, n\}$. Define $\widetilde{F}_i^{k} = F_i^k \setminus \{ X_i(r_{i}^k) \}$ and $\widetilde{B}_i^{k} = B_i^k \setminus \{X_i(l_i^k)\}$. We look at the individual terms separately:
\begin{align} \label{eq:Ycondexpec}
    Y_{g,i}^k(\widetilde{\alpha}) &= \widetilde{\mathbb{P}}(X_i(\tau_k-) = g \given \mathbb{X}_i^\mathrm{O}) \nonumber \\ 
    &= \widetilde{\mathbb{P}}(X_i(\tau_k-) = g \given \widetilde{B}^{k}_i, X_i(l_i^k) = a_i^k, X_i(r_i^k) = b_i^k,  \widetilde{F}^{k}_i) \nonumber \\
    &= \frac{  \widetilde{\mathbb{P}}(X_i(r_i^k) = b_i^k, \widetilde{F}^{k}_i\given X_i(\tau_k-) = g,   \widetilde{B}^{k}_i,  X_i(l_i^k) = a_i^k) \cdot  \widetilde{\mathbb{P}}(X_i(\tau_k-) = g \given \widetilde{B}^{k}_i,  X_i(l_i^k) = a_i^k)}{\widetilde{\mathbb{P}}(X_i(r_i^k) = b_i^k, \widetilde{F}^{k}_i\given  X_i(l_i^k) = a_i^k,  \widetilde{B}^{k}_i )} \nonumber \\
    &\overset{Markov}= \frac{  \widetilde{\mathbb{P}}(X_i(r_i^k) = b_i^k, \widetilde{F}^{k}_i\given X_i(\tau_k-) = g) \cdot  \widetilde{\mathbb{P}}(X_i(\tau_k-) = g \given  X_i(l_i^k) = a_i^k)}{\widetilde{\mathbb{P}}(X_i(r_i^k) = b_i^k, \widetilde{F}^{k}_i\given  X_i(l_i^k) = a_i^k)} \nonumber \\
    &= \frac{  \widetilde{\mathbb{P}}(\widetilde{F}^{k}_i\given X_i(r_i^k) = b_i^k, X_i(\tau_k-) = g) \cdot \widetilde{\mathbb{P}}(X_i(r_i^k) = b_i^k \given X_i(\tau_k-) = g) \cdot  \widetilde{\mathbb{P}}(X_i(\tau_k-) = g \given  X_i(l_i^k) = a_i^k)}{\widetilde{\mathbb{P}}(\widetilde{F}^{k}_i \given X_i(r_i^k) = b_i^k, X_i(\tau_k-) = g) \cdot \widetilde{\mathbb{P}}(X_i(r_i^k) = b_i^k \given  X_i(l_i^k) = a_i^k)} \nonumber \\
    &\overset{Markov}= \frac{  \widetilde{\mathbb{P}}(X_i(r_i^k) = b_i^k \given X_i(\tau_k-) = g) \cdot  \widetilde{\mathbb{P}}(X_i(\tau_k-) = g  \given  X_i(l_i^k) = a_i^k)}{\widetilde{\mathbb{P}}(X_i(r_i^k) = b_i^k \given  X_i(l_i^k) = a_i^k)} \nonumber \\
    &= \frac{\widetilde{P}_{a_i^{k},g}(l_i^{k},\tau_{k}-) \cdot \widetilde{P}_{g,b_i^{k}}(\tau_{k}-,r_i^{k})}
                {\widetilde{P}_{a_i^{k},b_i^{k}}(l_i^{k},r_i^{k})},
\end{align} with $\widetilde{P}_{gh}(s, t)$ the $(g,h)-$th entry of the product integral over the current estimates $\widetilde{\mathbf{P}}(s, t) = \prodi_{s < u \leq t} \left( \mathbf{I} + \textrm{d}\widetilde{\mathbf{A}}(u) \right)$.
We continue to the next term:
\begin{align} \label{eq:dcondexpec}
    d_{gh,i}^k &= \widetilde{\mathbb{P}}(X_i(\tau_k-) = g, X_i(\tau_k) = h \given \mathbb{X}_i^\mathrm{O}) \nonumber \\
    &= \widetilde{\mathbb{P}}(X_i(\tau_k-) = g \given  \mathbb{X}_i^\mathrm{O}) \cdot \widetilde{\mathbb{P}}(X_i(\tau_k) = h  \given X_i(\tau_k-) = g, \mathbb{X}_i^\mathrm{O}).
\end{align} The first term is simply $Y_{g,i}^k(\widetilde{\alpha})$, so we take a further look at the second term:
\begin{align} \label{eq:dtermcondexpec}
    \widetilde{\mathbb{P}}(X_i(\tau_k) = h  \given X_i(\tau_k-) = g, \mathbb{X}_i^\mathrm{O}) &= \frac{\widetilde{\mathbb{P}}(X_i(\tau_k) = h, X_i(\tau_k-) = g, B^k_i, F^k_i)}{\widetilde{\mathbb{P}}(X_i(\tau_k-) = g, B^k_i, F^k_i)} \nonumber \\
    &= \frac{\widetilde{\mathbb{P}}(F^k  \given X_i(\tau_k) = h) \cdot \widetilde{\mathbb{P}}(X_i(\tau_k) = h  \given X_i(\tau_k-) = g)}{\widetilde{\mathbb{P}}(F^k  \given X_i(\tau_k-) = g)} \nonumber \\
    &= \frac{\widetilde{\mathbb{P}}(X_i(r_i^k) = b_i^k  \given X_i(\tau_k) = h) \cdot \widetilde{\alpha}_{gh}^k}{\widetilde{\mathbb{P}}(X_i(r_i^k) = b_i^k  \given X_i(\tau_k-) = g)} \nonumber \\
    &= \frac{\widetilde{\mathbb{P}}(X_i(r_i^k) = b_i^k  \given X_i(\tau_k) = h) \cdot \widetilde{\alpha}_{gh}^k}{\widetilde{\mathbb{P}}(X_i(r_i^k) = b_i^k  \given X_i(\tau_k-) = g)} \nonumber \\
    &= \frac{\widetilde{P}_{h,b_i^k}(\tau_k, r_i^k) \cdot \widetilde{\alpha}_{gh}^k}{\widetilde{P}_{g,b_i^k}(\tau_k-, r_i^k)}.
\end{align} As the cumulative intensities were assumed to be right-continuous step functions, we obtain that:
\begin{align*}
    \widetilde{P}_{gh}(\tau_k-, t) = \widetilde{P}_{gh}(\tau_{k-1}, t),
\end{align*} for any $k \in 1, \ldots, K$ and any $t > \tau_k$. Using this fact and substituting Equations \eqref{eq:dtermcondexpec} and \eqref{eq:Ycondexpec} into Equation \eqref{eq:dcondexpec} and summing over $n$ we obtain the final result.

\end{proof}

\begin{proposition} \label{prop:KKTconditions}
The conditional expected value of the complete data log-likelihood for interval-censored multi state data (see Equation \eqref{eq:condexpecll}) over the optimisation region:
\begin{align*}
    C_\alpha &= \left\lbrace \alpha_{gh}^k, g \neq h \in \mathcal{H}, k = 1, \ldots, K; \alpha_{gh}^k \geq 0, \sum_{h \neq g = 1}^H \alpha_{gh}^k \leq 1 \right\rbrace,
\end{align*}
is maximised by the following expression:
\begin{align*}
    \alpha_{gh}^k &= \begin{cases}
        \frac{d_{gh}^k(\widetilde{\alpha})}{Y_g^k(\widetilde{\alpha})}, & \mu_g^k = 0, \\
        \frac{d_{gh}^k(\widetilde{\alpha})}{\sum_{h \leftarrow g} d_{gh}^k(\widetilde{\alpha})}, & \mu_g^k > 0,
    \end{cases}
\end{align*} with $\mu_g^k = \max\left(  0,  \sum_{h \leftarrow g}d_{gh}^k(\widetilde{\alpha}) - Y_g^k(\widetilde{\alpha}) \right).$
\end{proposition}

\begin{proof}
Maximisation of the complete data log-likelihood function can be phrased as a convex optimization problem. The primal problem is given by (denoting $l^C(\alpha) = \widetilde{\mathbb{E}}[\ell^C]$):
\begin{align*}
\begin{array}{llll}
    & \max  &l^C(\alpha) = \sum_{k=1}^K \sum_{(g,h) \in \mathcal{V}} d_{gh}^k(\widetilde{\alpha}) \log(\alpha_{gh}^k) - \sum_{k=1}^K \sum_{(g,h) \in \mathcal{V}} \alpha_{gh}^k Y_{g}^k(\widetilde{\alpha}) &\\
    & \text{s.t.} &\alpha_{gh}^k \geq 0, & g\neq h \in \mathcal{H}, k \in \{1, \ldots, K\}, \\
    & & \sum_{h \neq g} \alpha_{gh}^k \leq 1, & g \in \mathcal{H}, k \in \{1, \ldots, K\},
\end{array}
\end{align*} or rewritten as a minimization problem:
\begin{align*}
    \begin{array}{llll}
    & \min  &-l^C(\alpha) =  \sum_{k=1}^K \sum_{(g,h) \in \mathcal{V}} \alpha_{gh}^k Y_{g}^k(\widetilde{\alpha}) - \sum_{k=1}^K \sum_{(g,h) \in \mathcal{V}} d_{gh}^k(\widetilde{\alpha}) \log(\alpha_{gh}^k) &\\
    & \text{s.t.} &-\alpha_{gh}^k \leq 0, & g\neq h \in \mathcal{H}, k \in \{1, \ldots, K\}, \\
    & & \sum_{h \neq g} \alpha_{gh}^k - 1 \leq 0, & g \in \mathcal{H}, k \in \{1, \ldots, K\}.
\end{array}
\end{align*} As the logarithm is a concave function, minus the logarithm is convex and the addition of linear terms results in a convex function $-l^C(\alpha)$. The Lagrangian is given by:
\begin{align*}
    \mathcal{L}(\alpha, \mu, \lambda) &= \sum_{k=1}^K \sum_{(g,h) \in \mathcal{V}} \alpha_{gh}^k Y_{g}^k(\widetilde{\alpha}) - \sum_{k=1}^K \sum_{(g,h) \in \mathcal{V}} d_{gh}^k(\widetilde{\alpha}) \log(\alpha_{gh}^k) \\ & - \sum_{g = 1}^H \sum_{h \neq g}^H \sum_{k=1}^K \lambda_{gh}^k \alpha_{gh}^k + \sum_{g=1}^H \sum_{k = 1}^K \mu_{g}^k \left(  \sum_{h\neq g = 1}^H \alpha_{gh}^k - 1\right),
\end{align*} with $\lambda_{gh}^k, \mu_g^k$ called Lagrange multipliers. A necessary and sufficient condition for $\alpha$ to be a (local) minimum to above minimisation problem is given by the Karush-Kuhn-Tucker (KKT) conditions (See Section 5.5.3 of \cite{Boyd_Vandenberghe_2023}):
\begin{align*}
\begin{array}{c c c c c}
(1) &  \frac{\partial \mathcal{L}(\alpha, \mu)}{\partial \alpha_{gh}^k} =  & Y_g^k - \frac{d_{gh}^k}{\alpha_{gh}^k}  - \lambda_{gh}^k + \mu_g^k &=  0,  & g \neq h \in \mathcal{H}, k \in \{1, \ldots, K\}, \\
(2) &   & \lambda_{gh}^k \alpha_{gh}^k &= 0,  & g \neq h \in \mathcal{H}, k \in \{1, \ldots, K\}, \\
(3) &  & \mu_g^k \left( \sum_{h \leftarrow g} \alpha_{gh}^k - 1 \right) &= 0,  & g \in \mathcal{H}, k \in \{1, \ldots, K\}, \\
(4) &   & \alpha_{gh}^k &\geq 0,  & g\neq  h \in \mathcal{H}, k \in \{1, \ldots, K\}, \\
(5) &   & \sum_{h \leftarrow g} \alpha_{gh}^k - 1  &\leq 0,  & g \in \mathcal{H}, k \in \{1, \ldots, K\}, \\
(6) &   & \lambda_{gh}^k, \mu_{g}^k  &\geq 0,  & g\neq h \in \mathcal{H}, k \in \{1, \ldots, K\}. \\
\end{array}
\end{align*} When $\sum_{h \neq g = 1}^H \alpha_{gh}^k - 1 < 0$, constraint $(3)$ is called inactive. When the constraint is inactive, we must have that $\mu_g^k = 0$. When the constraint is active, $\mu_g^k$ tells us how much we can improve the objective function by changing the right-hand side of the inequality constraint. Note that $\mu_g^k$ must be positive, as increasing the right hand side of inequality constraint $(5)$ can improve the objective function. 

From equality $(1)$ in the KKT conditions we recover:
\begin{align*}
    Y_g^k(\widetilde{\alpha}) - \frac{d_{gh}^k(\widetilde{\alpha})}{\alpha_{gh}^k} - \lambda_{gh}^k + \mu_g^k = 0.
\end{align*} Note that this is undefined for $\alpha_{gh}^k = 0$. Clearly, if $d_{gh}^k(\widetilde{\alpha}) = 0$, then such a transition is not possible under our current estimate of $\alpha$, so we define $\frac{0}{0} = 0$. Suppose that $\alpha_{gh}^k > 0$. In that case, the complementary slackness condition $\lambda_{gh}^k \alpha_{gh}^k = 0$ ensures that $\lambda_{gh}^k = 0$. We obtain:
\begin{align} \label{eq:KKT1}
    Y_g^k(\widetilde{\alpha}) - \frac{d_{gh}^k(\widetilde{\alpha})}{\alpha_{gh}^k} +  \mu_g^k = 0.
\end{align} Multiply both sides by $\alpha_{gh}^k > 0$:
\begin{align*}
    \alpha_{gh}^k Y_g^k(\widetilde{\alpha}) - d_{gh}^k(\widetilde{\alpha}) + \alpha_{gh}^k \mu_g^k = 0.
\end{align*} Sum over all states $h$ directly connected to $g$:
\begin{align*}
    Y_g^k(\widetilde{\alpha}) \sum_{h \leftarrow g} \alpha_{gh}^k - \sum_{h \leftarrow g} d_{gh}^k(\widetilde{\alpha}) + \mu_g^k \sum_{h \leftarrow g}\alpha_{gh}^k  = 0.
\end{align*} We add and subtract $\mu_g^k$ to obtain:
\begin{align*}
    Y_g^k(\widetilde{\alpha}) \sum_{h \leftarrow g} \alpha_{gh}^k - \sum_{h \leftarrow g} d_{gh}^k(\widetilde{\alpha}) + \mu_g^k \left( \sum_{h \leftarrow g}\alpha_{gh}^k - 1 \right) + \mu_g^k = 0,
\end{align*} and using KKT Condition (3):
\begin{align*} 
     \mu_g^k = \sum_{h \leftarrow g} d_{gh}^k(\widetilde{\alpha}) -  Y_g^k(\widetilde{\alpha}) \sum_{h \leftarrow g} \alpha_{gh}^k.
\end{align*}  Finally, we obtain:
\begin{align} \label{eq:mugksingle}
    \sum_{h \leftarrow g} \alpha_{gh}^k  = \frac{- \mu_g^k + \sum_{h \leftarrow g} d_{gh}^k(\widetilde{\alpha})}{Y_g^k(\widetilde{\alpha}) }.
\end{align}

Alternatively, we rewrite Equation \eqref{eq:KKT1} to obtain:
\begin{align} \label{eq:alphaghsingle}
    \alpha_{gh}^k = \frac{d_{gh}^k(\widetilde{\alpha})}{Y_g^k(\widetilde{\alpha}) + \mu_g^k},
\end{align} and summing over $h \leftarrow g$:
\begin{align} \label{eq:alphaghksummed}
    \sum_{h \leftarrow g} \alpha_{gh}^k =  \frac{\sum_{h \leftarrow g} d_{gh}^k(\widetilde{\alpha})}{Y_g^k(\widetilde{\alpha}) + \mu_g^k}.
\end{align} Now equating Equations \eqref{eq:mugksingle} and \eqref{eq:alphaghksummed} we obtain:
\begin{align*}
    \frac{- \mu_g^k + \sum_{h \leftarrow g} d_{gh}^k(\widetilde{\alpha})}{Y_g^k(\widetilde{\alpha}) } = \frac{\sum_{h \leftarrow g} d_{gh}^k(\widetilde{\alpha})}{Y_g^k(\widetilde{\alpha}) + \mu_g^k}.
\end{align*} After some algebra we obtain:
\begin{align*}
    \mu_g^k \left(  \mu_g^k + Y_g^k(\widetilde{\alpha}) -  \sum_{h \leftarrow g} d_{gh}^k(\widetilde{\alpha}) \right) = 0.
\end{align*} Taking into account that $\mu_g^k \geq 0$ for a feasible solution, we must have:
\begin{align}\label{eq:mugksolutionmax}
    \mu_g^k = \max\left(  0,  \sum_{h \leftarrow g}d_{gh}^k(\widetilde{\alpha}) - Y_g^k(\widetilde{\alpha}) \right).
\end{align} Substituting Equation \eqref{eq:mugksolutionmax} into Equation \eqref{eq:alphaghsingle} we obtain the final result:
\begin{align} \label{eq:alphaghksolution2}  
    \alpha_{gh}^k &= 
    \begin{cases}
        \frac{d_{gh}^k(\widetilde{\alpha})}{Y_g^k(\widetilde{\alpha})}, & \mu_g^k = 0, \\
        \frac{d_{gh}^k(\widetilde{\alpha})}{\sum_{h \leftarrow g} d_{gh}^k(\widetilde{\alpha})}, & \mu_g^k > 0.
    \end{cases}
\end{align}

\end{proof}

\subsection{Interval censored multi state data with exactly observed states}
\label{sec:EMderivationsPanelExact}

\begin{proposition} \label{prop:condexpecexact}
Consider the conditional expectation of the complete data log-likelihood function for interval-censored multi state data:
\begin{align} \label{eq:condexpecllexact}
    \mathbb{E}[ \ell^\mathrm{C}  \given \mathrm{O}, \widetilde{\alpha}] &= \sum_{k=1}^K \sum_{(g,h) \in \mathcal{V}} \left\lbrace \mathbb{E}\left[d_{gh}^k   \given \mathrm{O}, \widetilde{\alpha} \right] \log(\alpha_{gh}^k) -  \alpha_{gh}^k \mathbb{E}\left[Y_{g}^k   \given \mathrm{O}, \widetilde{\alpha}  \right] \right\rbrace.
\end{align} The conditional expectations for interval-censored multi state data with transition times exactly known for a subset of states $\mathcal{E} \subseteq \mathcal{H}$ are given by:
\begin{align*}
    d_{gh,i}^k(\widetilde{\alpha}) \coloneqq \mathbb{E}\left[d_{gh,i}^k   \given \mathrm{O}, \widetilde{\alpha} \right] &= 
    \begin{cases}
        \frac{\widetilde{P}_{a_i^{k},g}(l_i^{k},\tau_{k-1}) \cdot \widetilde{\alpha}_{gh}^k \cdot \widetilde{P}_{h,b_i^{k}}(\tau_k,r_i^{k})}
    {\widetilde{P}_{a_i^{k},b_i^{k}}(l_i^{k},r_i^{k})}, & \text{if } b_i^k \notin \mathcal{E} \\
    \frac{\widetilde{P}_{a_i^kg}(l_i^k, \tau_{k-1}) \cdot \widetilde{\alpha}_{gh}^k \cdot \sum_{m \in \mathcal{R}_{b_{i}^k}} \widetilde{\alpha}_{mb_i^k}^{k_{r_i}} \widetilde{P}_{hm}(\tau_k, \tau_{k_{r_i}-1}) }{\sum_{m \in \mathcal{R}_{b_{i}^k}} \widetilde{\alpha}_{mb_{i}^k}^{k_{r_i}} \widetilde{P}_{a_i^km}(l_i^k, \tau_{k_{r_i}-1})}, & \text{if } b_i^k \in \mathcal{E}\\
    \end{cases} \\
    Y_{g,i}^k(\widetilde{\alpha}) \coloneqq \mathbb{E}\left[Y_{g,i}^k   \given \mathrm{O}, \widetilde{\alpha}  \right] &= 
    \begin{cases}
        \frac{\widetilde{P}_{a_i^{k},g}(l_i^{k},\tau_{k-1}) \cdot \widetilde{P}_{g,b_i^{k}}(\tau_{k}-,r_i^{k})}{\widetilde{P}_{a_i^{k},b_i^{k}}(l_i^{k},r_i^{k})}, & \text{if } b_i^k \notin \mathcal{E} \\
        \frac{\widetilde{P}_{a_i^kg}(l_i^k, \tau_{k-1}) \cdot   \sum_{m \in \mathcal{R}_{b_{i}^k}} \widetilde{\alpha}_{mb_{i}^k}^{k_{r_i}} \widetilde{P}_{gm}(\tau_{k-1}, \tau_{k_{r_i}-1})  }{\sum_{m \in \mathcal{R}_{b_{i}^k}} \widetilde{\alpha}_{mb_{i}^k}^{k_{r_i}} \widetilde{P}_{a_i^km}(l_i^k, \tau_{k_{r_i}-1})}, & \text{if } b_i^k \in \mathcal{E}
    \end{cases}
\end{align*} with $\tau_{k_{r_i}} \coloneqq r_i^k$, $d_{gh}^k(\widetilde{\alpha}) = \sum_{i=1}^n d_{gh,i}^k(\widetilde{\alpha})$, $Y_{g}^k(\widetilde{\alpha}) = \sum_{i=1}^n Y_{g,i}^k(\widetilde{\alpha})$ and  $\widetilde{\mathbf{P}}(s, t) = \prodi_{s < u \leq t} \left( \mathbf{I} + \textrm{d}\widetilde{\mathbf{A}}(u) \right)$ the product integral of the current estimates of the cumulative transition intensities.
\end{proposition}

\begin{proof}
We follow the beginning of the proof of Proposition \ref{prop:condexpecinterval}. Fix $k$ and $i$ and consider the situation when $b_i^k \notin \mathcal{E}$. It is apparent that the result from Proposition \ref{prop:condexpecinterval} still holds, as no additional information is available in that case. We therefore consider the case when $b_i^k \in \mathcal{E}$ and look at the individual terms separately:
\begin{align*}
    Y_{g,i}^k(\widetilde{\alpha}) &= \widetilde{\mathbb{P}}(X_i(\tau_k-) = g  \given \mathbb{X}_i^\mathrm{O}) \\
    &= \widetilde{\mathbb{P}}(X_i(\tau_k-) = g  \given X_i(r_i^k) = b_i^k, \widetilde{B}^{k}_i, \widetilde{F}^{k}_i, X_i(l_i^k) = a_i^k, b_i^k \in \mathcal{E}) \\
    &= \frac{  \widetilde{\mathbb{P}}(\widetilde{F}^{k}, X_i(r_i^k) = b_i^k  \given X_i(\tau_k-) = g,   \widetilde{B}^{k}_i,  X_i(l_i^k) = a_i^k, b_i^k \in \mathcal{E}) \cdot  \widetilde{\mathbb{P}}(X_i(\tau_k-) = g  \given  X_i(l_i^k) = a_i^k, \widetilde{B}^{k}_i, b_i^k \in \mathcal{E})}{\widetilde{\mathbb{P}}(\widetilde{F}^{k}, X_i(r_i^k) = b_i^k \given  X_i(l_i^k) = a_i^k,  \widetilde{B}^{k}_i, b_i^k \in \mathcal{E})} \\
    &\overset{Markov}= \frac{  \widetilde{\mathbb{P}}(\widetilde{F}^{k}, X_i(r_i^k) = b_i^k \given X_i(\tau_k-) = g, b_i^k \in \mathcal{E}) \cdot  \widetilde{\mathbb{P}}(X_i(\tau_k-) = g  \given  X_i(l_i^k) = a_i^k, b_i^k \in \mathcal{E})}{\widetilde{\mathbb{P}}(\widetilde{F}^{k}, X_i(r_i^k) = b_i^k \given  X_i(l_i^k) = a_i^k,  b_i^k \in \mathcal{E})} \\
    &= \frac{  \widetilde{\mathbb{P}}(\widetilde{F}^{k}, X_i(r_i^k) = b_i^k \given X_i(\tau_k-) = g, b_i^k \in \mathcal{E}) \cdot  \widetilde{\mathbb{P}}(X_i(\tau_k-) = g  \given  X_i(l_i^k) = a_i^k)}{\widetilde{\mathbb{P}}(\widetilde{F}^{k}, X_i(r_i^k) = b_i^k \given  X_i(l_i^k) = a_i^k, b_i^k \in \mathcal{E})} \\
    &= \frac{  \widetilde{\mathbb{P}}(X_i(r_i^k) = b_i^k \given X_i(\tau_k-) = g, b_i^k \in \mathcal{E}) \cdot  \widetilde{\mathbb{P}}(X_i(\tau_k-) = g  \given  X_i(l_i^k) = a_i^k)}{\widetilde{\mathbb{P}}(X_i(r_i^k) = b_i^k \given  X_i(l_i^k) = a_i^k, b_i^k \in \mathcal{E})}.
\end{align*}
$\mathbb{X}_i^\mathrm{O}$ is the observed information for subject $i$ as defined in Section \ref{sec:methodsdata}. As $b_i^k \in \mathcal{E}$, the subject must have been in $\mathcal{R}_{b_i^k}$ at time $r_i^k-$. We therefore define $k_{r_i}$ such that $\tau_{k_{r_i}} = r_i^k$ and condition on time $\tau_{k_{r_i}-1}$ using the law of total probability in the numerator above:
 \begin{align*}
     \widetilde{\mathbb{P}}(X_i(r_i^k) = b_i^k \given X_i(\tau_k-) = g, b_i^k \in \mathcal{E}) &= \sum_{m \in \mathcal{R}_{b_{i}^k}} \widetilde{\mathbb{P}}(X_i(r_i^k) = b_i^k \given X_i(\tau_k-) = g, X_i(r_i^k-) = m, b_i^k \in \mathcal{E}) \cdot \\ & \hspace{5em}\widetilde{\mathbb{P}}(X_i(r_i^k-) = m  \given  X_i(\tau_k-) = g, b_i^k \in \mathcal{E}) \\
    &= \sum_{m \in \mathcal{R}_{b_{i}^k}} \widetilde{\mathbb{P}}(X_i(r_i^k) = b_i^k \given X_i(r_i^k-) = m, b_i^k \in \mathcal{E}) \cdot \\ & \hspace{5em}\widetilde{\mathbb{P}}(X_i(r_i^k-) = m  \given  X_i(\tau_k-) = g) \\
    &= \sum_{m \in \mathcal{R}_{b_{i}^k}} \widetilde{P}_{mb_i^k}(r_i^k-, r_i^k) \widetilde{P}_{gm}(\tau_{k-1}, r_i^k-) \\
    &= \sum_{m \in \mathcal{R}_{b_{i}^k}} \widetilde{\alpha}_{mb_{i}^k}^{k_{r_i}} \widetilde{P}_{gm}(\tau_{k-1}, \tau_{k_{r_i}-1}).
 \end{align*} Similarly, we obtain for the denominator:
\begin{align*}
    \widetilde{\mathbb{P}}(X_i(r_i^k) = b_i^k \given  X_i(l_i^k) = a_i^k, b_i^k \in \mathcal{E}) &= \sum_{m \in \mathcal{R}_{b_{i}^k}} \widetilde{\alpha}_{mb_{i}^k}^{k_{r_i}} \widetilde{P}_{a_i^km}(l_i^k, \tau_{k_{r_i}-1}).
\end{align*} Substituting these back into the original expression we obtain:
\begin{align}\label{eq:Y3}
    Y_{g,i}^k(\widetilde{\alpha}) &= \frac{\widetilde{P}_{a_i^kg}(l_i^k, \tau_{k-1}) \cdot   \sum_{m \in \mathcal{R}_{b_{i}^k}} \widetilde{\alpha}_{mb_{i}^k}^{k_{r_i}} \widetilde{P}_{gm}(\tau_{k-1}, \tau_{k_{r_i}-1})  }{\sum_{m \in \mathcal{R}_{b_{i}^k}} \widetilde{\alpha}_{mb_{i}^k}^{k_{r_i}} \widetilde{P}_{a_i^km}(l_i^k, \tau_{k_{r_i}-1})}.
\end{align}  

We continue on to:
\begin{align*}
    d_{gh,i}^k(\widetilde{\alpha}) &= \widetilde{\mathbb{P}}(X_i(\tau_k-) = g, X_i(\tau_k) = h \given \mathbb{X}_i^\mathrm{O}) \\
    &= \widetilde{\mathbb{P}}(X_i(\tau_k-) = g \given  \mathbb{X}_i^\mathrm{O}) \cdot \widetilde{\mathbb{P}}(X_i(\tau_k) = h  \given X_i(\tau_k-) = g, \mathbb{X}_i^\mathrm{O}),
\end{align*} and again focus only on the second term:
\begin{align*}
    \widetilde{\mathbb{P}}(X_i(\tau_k) = h  \given X_i(\tau_k-) = g, \mathbb{X}_i^\mathrm{O}) &= \frac{\widetilde{\mathbb{P}}(F^k  \given X_i(\tau_k) = h, b_i^k \in \mathcal{E}) \cdot \widetilde{\mathbb{P}}(X_i(\tau_k) = h  \given X_i(\tau_k-) = g, b_i^k \in \mathcal{E})}{\widetilde{\mathbb{P}}(F^k  \given X_i(\tau_k-) = g, b_i^k \in \mathcal{E})} \\
    &= \frac{\widetilde{\mathbb{P}}(X_i(r_i^k) = b_i^k  \given X_i(\tau_k) = h, b_i^k \in \mathcal{E}) \cdot \widetilde{\alpha}_{gh}^k}{\widetilde{\mathbb{P}}(X_i(r_i^k) = b_i^k  \given X_i(\tau_k-) = g, b_i^k \in \mathcal{E})}.
\end{align*} The term in the denominator was already expanded above, so we focus on the term in the numerator (in a similar way as before):
\begin{align*}
    \widetilde{\mathbb{P}}(X_i(r_i^k) = b_i^k  \given X_i(\tau_k) = h, b_i^k \in \mathcal{E}) &= \sum_{m \in \mathcal{R}_{b_{i}^k}} \widetilde{\alpha}_{mb_i^k}^{k_{r_i}} \widetilde{P}_{hm}(\tau_k, \tau_{k_{r_i}-1}). 
\end{align*} Substituting these results into the original expression we obtain:
\begin{align}\label{eq:d3}
    d_{gh,i}^k &= \frac{\widetilde{P}_{a_i^kg}(l_i^k, \tau_{k-1}) \cdot \widetilde{\alpha}_{gh}^k \cdot \sum_{m \in \mathcal{R}_{b_{i}^k}} \widetilde{\alpha}_{mb_i^k}^{k_{r_i}} \widetilde{P}_{hm}(\tau_k, \tau_{k_{r_i}-1}) }{\sum_{m \in \mathcal{R}_{b_{i}^k}} \widetilde{\alpha}_{mb_{i}^k}^{k_{r_i}} \widetilde{P}_{a_i^km}(l_i^k, \tau_{k_{r_i}-1})}.
\end{align}
\end{proof}

\section{Latent Poisson Expectation Maximisation algorithm}
\label{sec:EMderivationslatentpoisson}

In this section, we loosely describe the EM algorithm for the NPMLE of interval-censored multi state data based on latent Poisson variables \cite{Gu2023}. This Section therefore does not contain any original work.

Their EM algorithm is very similar to the one described in Section \ref{sec:EMderivationsPanel}, so most of the previous notation carries over. We do however need to introduce some new notation first, mainly for the latent Poisson variables. We consider the interval-censored multi state observed-data likelihood. Similar to above, the cumulative intensity function is assumed to be a right-continuous step function with jumps only at the unique observation times.

Fix $i$ and $j$, then $[t_{i,j-1}, t_{ij}]$ is some observation interval with observed states $x_{i,j-1}$ and $x_{ij}$. Let $w$ be the value such that $\tau_{w-1} = t_{i,j-1}$, and $q$ the value such that $\tau_{w+q+1} = t_{ij}$. Then $t_{i,j-1} = \tau_{w-1} < \tau_{w} < \ldots < \tau_{w+q} < \tau_{w+q+1} = t_{ij}$ is a grid spanning the observation interval. A transition from $x_{i,j-1}$ to $x_{ij}$ in the observation interval must have happened through one of the possible transition paths $(u_{w-1}, u_{w}, \ldots, u_{iw+q}, u_{w+q+1})$ with $u_{w}, \ldots, u_{w+q}$ the unobserved states at the corresponding times $\tau_{w}, \ldots, \tau_{w+q}$. For such a possible transition path, they define the event $V_i(u_{w}, \ldots, u_{w+q}, t_{i,j-1}, t_{ij}, x_{i,j-1}, x_{ij})$ through latent Poisson random variables $W_{gh,i}^k$ as follows. For $k = w, \ldots, w + q + 1$, if $u_{k-1} \neq u_k$ then $W_{u_{k-1}u_k, i}^k > 0$ and $W_{u_{k-1} u', i}^k = 0$ for all $u' \neq u_{k-1}, u_k$. Else if $u_{k-1} = u_k$ we have that $W_{u_{k-1}u', i}^k = 0$ for all $u' \neq u_{k-1}$. They then define the event $Y_i(t_{i,j-1}, t_{ij}, x_{i,j-1}, x_{ij}) = \bigcup_{(u_w, \ldots, u_{w+q}) \in \mathcal{A}_{w+q}} V_i(u_{w}, \ldots, u_{w+q}, t_{i,j-1}, t_{ij}, x_{i,j-1}, x_{ij})$ with $\mathcal{A}_{w+q}$ the set of possible transitions connecting $x_{i,j-1}$ and $x_{ij}$. They show that maximising the interval-censored multi state observed-data likelihood is the same as maximising the likelihood based on the observations $\mathcal{O}_i = \bigcap_{j=1}^{n_i} Y_i(t_{i,j-1}, t_{ij}, x_{i,j-1}, x_{ij})$.

They show that the complete-data log likelihood for these latent Poisson variables is given by:
\begin{align}
    \label{eq:completedatalikpoisson2}
    \ell^{\mathrm{P}} &= \sum_{i=1}^n \left(  \sum_{k=1}^K \sum_{(g,h) \in \mathcal{V}} \mathds{1}\{ \tau_k \leq t_{in_i} \} \left[ W_{gh,i}^k \log(\alpha_{gh}^k) - \alpha_{gh}^k - \log(W_{gh,i}^k!)  \right]  \right).
\end{align} This likelihood is quite different compared to the complete-data log likelihood based on the multinomial distribution considered by us. 

To determine an update rule for $\alpha_{gh}^k$, they calculate the conditional expectation of the Poisson variables given the observed data and current estimates of the cumulative transition intensities:
\begin{align}
    \label{eq:expectedpoissonvariables}
    \widetilde{\mathbb{E}}[W_{gh, i}^k  \given O] = \mathbb{E}[W_{gh, i}^k  \given O, \widetilde{\alpha}]  &= \frac{\sum_{g' \neq g} \widetilde{P}_{a_i^k g'}(l_i^k, \tau_k-) \widetilde{P}_{g'b_i^k}(\tau_k-, r_i^k)}{\widetilde{P}_{a_i^k b_i^k}(l_i^k, r_i^k)} \widetilde{\alpha}_{gh}^k \\ &+ \frac{ \widetilde{P}_{a_i^k g}(l_i^k, \tau_k-) \widetilde{\alpha}_{gh}^k \widetilde{P}_{hb_i^k}(\tau_k, r_i^k)}{\widetilde{P}_{a_i^k b_i^k}(l_i^k, r_i^k)} \exp \left(-\sum_{h' \leftarrow g, h' \neq h} \widetilde{\alpha}_{gh'}^k\right).
\end{align}

The update rule (M-step) in the EM algorithm is then given by:
\begin{align}
    \label{eq:poissonemupdaterule}
    \alpha_{gh}^k &= \frac{\sum_{i=1}^n \mathds{1}\{\tau_k \leq t_{i,n_i}\} \widetilde{\mathbb{E}}[W_{gh,i}^k  \given O] }{\sum_{i=1}^n \mathds{1}\{\tau_k \leq t_{i,n_i}\}}.
\end{align}

Contrary to our approach in Section \ref{sec:EMderivationsPanel} they do not consider the KKT conditions to make sure that (sum of the) updated jumps in the intensities is bounded by zero and one in the M-step. In practice, this is unlikely to be an issue in non-parametric estimation, but might be problematic when covariates are included.

\section{Latent Poisson EM - initial estimate dependence}
\label{sec:poissoneminitialconditions}

In this section, we show that the initial intensity estimates for transitions out of non-absorbing states that are not covered by the initial state cannot be changed by the latent Poisson approach.

Consider $g, h \in \mathcal{H}$ such that the transition $g \to h$ is possible. Assume that no subject starts at time $0$ in state $g$. Consider the first bin $[0, \tau_1]$ and a single subject $i \in \{1, \ldots, n\}$. The contribution of this subject to the value of $\alpha_{gh}^1$ is given by Equation \eqref{eq:expectedpoissonvariables}:
\begin{align*}
    \widetilde{E}[W_{gh,i}^1  \given O]  &= \frac{\sum_{g' \neq g} \widetilde{P}_{a_i^1 g'}(l_i^1, \tau_1-) \widetilde{P}_{g'b_i^1}(\tau_1-, r_i^1)}{\widetilde{P}_{a_i^1 b_i^1}(l_i^1, r_i^1)} \widetilde{\alpha}_{gh}^1 \\ &+ \frac{ \widetilde{P}_{a_i^1 g}(l_i^1, \tau_1-) \widetilde{\alpha}_{gh}^1 \widetilde{P}_{hb_i^1}(\tau_1, r_i^1)}{\widetilde{P}_{a_i^1 b_i^1}(l_i^1, r_i^1)} \exp \left(-\sum_{h' \leftarrow g, h' \neq h} \widetilde{\alpha}_{gh'}^1\right).
\end{align*} Clearly $l_i^1 = \tau_1 - = 0$, so we obtain:
\begin{align*}
    \widetilde{E}[W_{gh,i}^1  \given O]  &= \frac{\sum_{g' \neq g} \widetilde{P}_{a_i^1 g'}(0, 0) \widetilde{P}_{g'b_i^1}(0, r_i^1)}{\widetilde{P}_{a_i^1 b_i^1}(0, r_i^1)} \widetilde{\alpha}_{gh}^1 \\ &+ \frac{ \widetilde{P}_{a_i^1 g}(0, 0) \widetilde{\alpha}_{gh}^1 \widetilde{P}_{hb_i^1}(\tau_1, r_i^1)}{\widetilde{P}_{a_i^1 b_i^1}(0, r_i^1)} \exp \left(-\sum_{h' \leftarrow g, h' \neq h} \widetilde{\alpha}_{gh'}^1\right).
\end{align*} Note that $\widetilde{P}_{a_i^1 g'}(0, 0)$ can only be non-zero when $g' = a_i^1$. The summation in the numerator of the first term therefore only yields a positive contribution when $g' = a_i^1$. As we assumed that no subjects start in state $g$, we must have that $g \neq a_i^1$ and therefore the second term is $0$. We obtain:
\begin{align*}
    \widetilde{E}[W_{gh,i}^1  \given O]  &= \frac{\sum_{g' \neq g} \widetilde{P}_{a_i^1 g'}(0, 0) \widetilde{P}_{g'b_i^1}(0, r_i^1)}{\widetilde{P}_{a_i^1 b_i^1}(0, r_i^1)} \widetilde{\alpha}_{gh}^1 = \frac{ \widetilde{P}_{a_i^1 a_i^1}(0, 0) \widetilde{P}_{a_i^1 b_i^1}(0, r_i^1)}{\widetilde{P}_{a_i^1 b_i^1}(0, r_i^1)} \widetilde{\alpha}_{gh}^1 = \widetilde{\alpha}_{gh}^1.
\end{align*} Therefore the contribution of all subjects is the same at each iteration. From Equation \eqref{eq:poissonemupdaterule} we then find that the estimate of $\widetilde{\alpha}_{gh}^1$ does not change over any iteration and the final estimate will simply be the initial estimate. The result shown here can also be extended to any $W_{gh,i}^k$ where no subject can be in state $g$ before time $\tau_k$. The summation in the numerator of the first term of Equation \eqref{eq:expectedpoissonvariables} represents the probability of reaching $b_i^k$ through any path that does not go through state $g$ at time $\tau_k-$. The denominator represents the probability of reaching $b_i^k$ through any path. As we assume that no subject can be in state $g$ at time $\tau_k-$, the numerator and denominator are equal and therefore the first term simply becomes $\widetilde{\alpha}_{gh}^k$. The second term represents the probability of making the $g \to h$ transition in the bin $(\tau_k-, \tau_k]$, but as we assumed no subject can be in state $g$ at time $\tau_k-$ this probability clearly is zero. This means that the estimate for the jumps in the transition intensities cannot be changed for an interval where there is zero probability to be in state $g$ at the start of that interval.


\section{Alternative estimation procedures}
\label{sec:alternativeestimationprocedures}

In this section, we discuss some alternative estimation procedures that can be used. The Nelson-Aalen estimator can be derived by using three different extensions of the likelihood function (see Section IV.1.5 of \cite{Andersen1997}). It is also possible to consider different extensions of the likelihood in the multi-state Markov setting, with the estimation procedure proposed in Section \ref{sec:modelcompletedatalik} following a ``Poisson'' approximation of the canonical multinomial extension described below.

\subsection{Canonical multinomial extension}
\label{sec:canonicalmultextension}

The canonical extension of the univariate survival model for multi-state Markov processes yields the following likelihood contribution per individual (See Section IV.4.1.5 of Andersen et al.): \cite{Andersen_Ravn_2024}
\begin{align*}
    L_i^\mathrm{C} &= \Prodi_{t \leq a_i f_i} \prod_{g \in \mathcal{H}} \left\lbrace  \prod_{h \neq g} \left\lbrace \left( Y_{g,i}(t) dA_{gh}(t) \right)^{dN_{gh,i}(t)} \right\rbrace \left( 1 - \sum_{z \leftarrow g} dA_{gz}(t) \right)^{Y_{g,i}(t) - \sum_{z \leftarrow g}dN_{gz,i}(t)}  \right\rbrace.
\end{align*} This has the interpretation of conditionally independent multinomially distributed numbers of jumps from each state $g$, such that $(dN_{gh}(t): h \neq g) \sim \mathrm{Mult}(Y_g(t), dA_{gh}(t)): h \neq g)$ given $\mathcal{F}_{t-}$. Note that for maximum likelihood estimation in the non-parametric case, we cannot assume the cumulative intensities are absolutely continuous, and therefore it is unclear whether the intensities exist. \cite{Andersen1997} We therefore attempt to estimate the cumulative intensities directly, allowing the cumulative intensities to have both continuous and discrete components. Taking the logarithm we obtain:
\begin{align*}
    \ell_{i}^\mathrm{C} &= \int_0^{t^*_{if_i}}  \sum_{g \in \mathcal{H}} \left\lbrace \sum_{h \neq g} dN_{gh,i}(t) \log(Y_{g,i}(t)dA_{gh}(t)) \right. \\ 
    &\left. + \left( Y_{g,i}(t) - \sum_{z \leftarrow g} dN_{gz,i}(t) \right) \log\left(1 - \sum_{z \leftarrow g}dA_{gz}(t)\right) \right\rbrace.
\end{align*}
Following the discussion in Section \ref{sec:modelcompletedatalik}, we maximize over the class of cumulative intensity functions with jumps only on the observed event times, reducing the integrals to summations. Clearly, $dN_{gh}(t) = 0$ whenever $(g,h) \notin \mathcal{V}$, therefore we can replace $\sum_{g \in \mathcal{H}} \sum_{h \neq g}$ with $\sum_{(g,h) \in \mathcal{V}}$:
 \begin{align*}
    \ell_{i}^\mathrm{C} &= \sum_{k=1}^K \left\lbrace \sum_{(g,h) \in \mathcal{V}}  d_{gh,i}^k \log(\alpha_{gh}^k) + \sum_{g \in \mathcal{H}} \left( Y_{g,i}^k - \sum_{z \leftarrow g} d_{gz,i}^k \right) \log\left(1 - \sum_{z \leftarrow g}\alpha_{gz}^k\right)  \right\rbrace.
\end{align*}
For the E-step (see Section \ref{sec:modelcensestep}), we calculate the expectation of the log-likelihood given the observed information, transforming the $Y'$s and $d'$s into expected quantities (notationally introducing tildes above the letters, see Equation \eqref{eq:paneldexpec}). This yields the following for the expected value of the complete data log likelihood:
\begin{align*}
    \mathbb{E}[\ell^\mathrm{C} \given \mathrm{O}, \widetilde{\alpha}] = \sum_{k=1}^K \left\lbrace \sum_{(g,h) \in \mathcal{V}}  d_{gh}^k(\widetilde{\alpha}) \log(\alpha_{gh}^k) + \sum_{g \in \mathcal{H}} \left( Y_{g}^k(\widetilde{\alpha}) - \sum_{z \leftarrow g} d_{gz}^k(\widetilde{\alpha}) \right) \log\left(1 - \sum_{z \leftarrow g}\alpha_{gz}^k\right)  \right\rbrace.
\end{align*}For the M-step we take the derivative to obtain the score function:
\begin{align*}
    \frac{d\mathbb{E}[\ell^\mathrm{C} \given \mathrm{O}, \widetilde{\alpha}]}{d\alpha_{gh}^k} &= \frac{d_{gh}^k(\widetilde{\alpha})}{\alpha_{gh}^k} - \frac{Y_{g}^k(\widetilde{\alpha}) - \sum_{z \leftarrow g}d_{gz}^k(\widetilde{\alpha})}{1 - \sum_{z \leftarrow g} \alpha_{gz}^k}.
\end{align*} Equating to zero and factoring out $\alpha_{gh}^k$ we obtain:
\begin{align}
    \alpha_{gh}^k &= \frac{d_{gh}^k(\widetilde{\alpha}) \left(1 - \sum_{z \leftarrow g}\alpha_{gz}^k + \alpha_{gh}^k  \right)}{Y_g^k(\widetilde{\alpha}) - \sum_{z \leftarrow g} d_{gz}^k(\widetilde{\alpha}) + d_{gh}^k(\widetilde{\alpha})}.
    \label{eq:canonicalmultinomialalphagh}
\end{align} Note that under the canonical multinomial extension the update rule for $\alpha_{gh}^k$ also depends on the values of jumps of the cumulative intensities for other transitions in the $k$-th bin which have not been estimated yet, leading to a system of equations. For notational convenience, define:
\begin{align*}
    Q_{gh}^k = \frac{d_{gh}^k(\widetilde{\alpha})}{Y_g^k(\widetilde{\alpha}) - \sum_{z \leftarrow g} d_{gz}^k(\widetilde{\alpha}) + d_{gh}^k(\widetilde{\alpha})}.
\end{align*}

Fix $g \in \mathcal{H}$ and $k \in \{1, \ldots, K\}$ and consider the column vectors $\vec{\alpha}_{g}^k = \left[ \alpha_{gh}^k \right]_{\{h \leftarrow g\}}$ and $\vec{Q}_{g}^k = \left[ Q_{gh}^k \right]_{\{h \leftarrow g\}}$. Assume without loss of generality that the first and last state can be reached from state $g$ in a single transition. Define:
\begin{align*}
    \vec{D}_g^k \coloneqq \mathrm{diag} \left( \vec{Q}_{g}^k   \right) = \left(  
    \begin{array}{c c c }
      \frac{d_{g1}^k(\widetilde{\alpha})}{Y_g^k(\widetilde{\alpha}) - \sum_{z \leftarrow g} d_{gz}^k(\widetilde{\alpha}) + d_{g1}^k(\widetilde{\alpha})}   &  \ldots & 0 \\
        \vdots & \ddots & \vdots \\
        0 & \ldots & \frac{d_{gH}^k(\widetilde{\alpha})}{Y_g^k(\widetilde{\alpha}) - \sum_{z \leftarrow g} d_{gz}^k(\widetilde{\alpha}) + d_{gH}^k(\widetilde{\alpha})} 
    \end{array} \right),
\end{align*}  as well as the hollow matrix:
\begin{align*}
    \vec{M}_g &\coloneqq \left( \begin{array}{cccc}
       0  & 1 & \ldots & 1 \\
        1 & 0 & \ldots & 1 \\
        \vdots &  & \ddots & \vdots \\
        1 & 1 & \ldots & 0
    \end{array} \right),
\end{align*}  with both matrices square with dimension equal to the number of state reachable from $g$: $|\{z \leftarrow g \}|$. Then we can write the system of equations as follows:
\begin{align*}
    \vec{\alpha}_g^k &= \vec{D}_g^k \left( \vec{1} - \vec{M}_g \vec{\alpha}_g^k  \right),
\end{align*} with $\vec{1}$ a column vector of ones. This equation can be solved by matrix inversion:
\begin{align}
    \vec{\alpha}_g^k &= \left( \vec{I} + \vec{D}_g^k \vec{M}_g \right)^{-1} \vec{D}_g^k \vec{1}.
    \label{eq:canonicalmultsolution}
\end{align} The matrix $\left( \vec{I} + \vec{D}_g^k \vec{M}_g \right)$ has entries $1$ on the diagonal with the rest of the row filled with the same value:
\begin{align*}
    \vec{I} + \vec{D}_g^k \vec{M}_g &= \left(
    \begin{array}{c c c c}
         1 &  \frac{d_{g1}^k(\widetilde{\alpha})}{Y_g^k(\widetilde{\alpha}) - \sum_{z \leftarrow g} d_{gz}^k(\widetilde{\alpha}) + d_{g1}^k(\widetilde{\alpha})} & \cdots & \frac{d_{g1}^k(\widetilde{\alpha})}{Y_g^k(\widetilde{\alpha}) - \sum_{z \leftarrow g} d_{gz}^k(\widetilde{\alpha}) + d_{g1}^k(\widetilde{\alpha})} \\
         \frac{d_{g2}^k(\widetilde{\alpha})}{Y_g^k(\widetilde{\alpha}) - \sum_{z \leftarrow g} d_{gz}^k(\widetilde{\alpha}) + d_{g2}^k(\widetilde{\alpha})} & 1 & \cdots & \frac{d_{g2}^k(\widetilde{\alpha})}{Y_g^k(\widetilde{\alpha}) - \sum_{z \leftarrow g} d_{gz}^k(\widetilde{\alpha}) + d_{g2}^k(\widetilde{\alpha})} \\
         \vdots & \cdots & \ddots & \vdots \\
         \frac{d_{gH}^k(\widetilde{\alpha})}{Y_g^k(\widetilde{\alpha}) - \sum_{z \leftarrow g} d_{gz}^k(\widetilde{\alpha}) + d_{gH}^k(\widetilde{\alpha})} & \frac{d_{gH}^k(\widetilde{\alpha})}{Y_g^k(\widetilde{\alpha}) - \sum_{z \leftarrow g} d_{gz}^k(\widetilde{\alpha}) + d_{gH}^k(\widetilde{\alpha})} & \cdots & 1
    \end{array}
    \right).
\end{align*}

Note that the matrix $\left( I + \vec{D}_g^k \vec{M}_g \right)$ in general has full rank (and is therefore invertible). The matrix is clearly not invertible when $Y_{g}^k(\widetilde{\alpha}) = \sum_{s \leftarrow g} d_{gs}^k(\widetilde{\alpha})$, as then all entries in the matrix will be equal to $1$. This only happens when, according to the current estimates $\widetilde{\alpha}$, there is no probability of staying in state $g$ during $(\tau_k-, \tau_k]$.

\subsection{Multinoulli extension}

Another possible extension is given by Andersen \& Ravn,\cite{Andersen_Ravn_2024} yielding the following contribution to the likelihood per subject:
\begin{align*}
    L_i^\mathrm{C} &= \Prodi_{t \leq a_i f_i}  \left\lbrace  \prod_{(g,h) \in \mathcal{V}} \left\lbrace \left( Y_{g,i}(t) dA_{gh}(t) \right)^{dN_{gh,i}(t)} \right\rbrace \left( 1 - \sum_{(g,h) \in \mathcal{V}} dA_{gh}(t) \right)^{1 - \sum_{(g,h) \in \mathcal{V}}dN_{gh,i}(t)}  \right\rbrace,
\end{align*} which has the interpretation of a multinomially distributed jump at each time point, such that $\left(dN_{gh}(t): (g,h) \in \mathcal{V}\right) \sim \mathrm{Mult}(1, dA_{gh}(t): (g,h) \in \mathcal{V})$ given $\mathcal{F}_{t-}$. The multinomial distribution with a single draw is often referred to as a multinoulli distribution, therefore we will call this the ``multinoulli'' extension of the likelihood. 

Taking very similar steps as in Section \ref{sec:canonicalmultextension} we are led to the following system of equations as maximizer of the expected complete-data likelihood:
\begin{align*}
    \alpha_{gh}^k &= \frac{d_{gh}^k(\widetilde{\alpha}) \left(1 - \sum_{(r,s) \in \mathcal{V}}\alpha_{rs}^k + \alpha_{gh}^k  \right)}{n - \sum_{(r,s) \in \mathcal{V}} d_{rs}^k(\widetilde{\alpha}) + d_{gh}^k(\widetilde{\alpha})}.
\end{align*} Compared with Equation \eqref{eq:canonicalmultinomialalphagh}, this expression has an $n$ instead of $Y_{g}^k(\widetilde{\alpha})$ and the summation is taken over all possible transitions instead of all possible transitions out of a single state $g$. Clearly, $n$ is the expected number of subjects ``at risk'' of a transition at any time $t$, making the two approaches very similar. The multinomial approach considers each transition separately, whereas the multinoulli approach obtains estimates by considering all possible transitions together.

We can solve the resulting system of equations in a similar fashion. Define:
\begin{align*}
    Q_{gh}^{k'} = \frac{d_{gh}^k(\widetilde{\alpha})}{n - \sum_{(r,s) \in \mathcal{V}} d_{rs}^k(\widetilde{\alpha}) + d_{gh}^k(\widetilde{\alpha})}.
\end{align*}

Fix $k \in \{1, \ldots, K\}$ and consider the column vectors $\vec{\alpha}^k = \left[ \alpha_{gh}^k \right]_{\{(g,h) \in \mathcal{V}\}}$ and $\vec{Q}^{k} = \left[ Q_{gh}^{k'} \right]_{\{(g,h) \in \mathcal{V}\}}$. Assume without loss of generality that the second state can be reached from the first and the second to last from the last state in a single transition. Define:
\begin{align*}
    \vec{D}^{k} \coloneqq \mathrm{diag} \left( \vec{Q}^{k}   \right) = \left(  
    \begin{array}{c c c }
      \frac{d_{12}^k(\widetilde{\alpha})}{n - \sum_{(r,s) \in \mathcal{V}}d_{rs}^k(\widetilde{\alpha}) + d_{12}^k(\widetilde{\alpha})}   &  \ldots & 0 \\
        \vdots & \ddots & \vdots \\
        0 & \ldots & \frac{d_{H, H-1}^k(\widetilde{\alpha})}{n - \sum_{(r,s) \in \mathcal{V}}d_{rs}^k(\widetilde{\alpha}) + d_{H, H-1}^k(\widetilde{\alpha})} 
    \end{array} \right),
\end{align*}  as well as the hollow matrix:
\begin{align*}
    \vec{M} &\coloneqq \left( \begin{array}{cccc}
       0  & 1 & \ldots & 1 \\
        1 & 0 & \ldots & 1 \\
        \vdots &  & \ddots & \vdots \\
        1 & 1 & \ldots & 0
    \end{array} \right),
\end{align*}  with both matrices square with dimension equal to the number of possible transitions in the model: $|\{(g,h) \in \mathcal{V}\}|$. Then we can write the system of equations as follows:
\begin{align*}
    \vec{\alpha}^k &= \vec{D}^k \left( \vec{1} - \vec{M} \vec{\alpha}^k  \right),
\end{align*} with $\vec{1}$ a column vector of ones. This equation can be solved by matrix inversion:
\begin{align}
    \vec{\alpha}^k &= \left( \vec{I} + \vec{D}^k \vec{M} \right)^{-1} \vec{D}^k \vec{1}.
    \label{eq:multinoullisolution}
\end{align} The matrix $\left( \vec{I} + \vec{D}^k \vec{M} \right)$ has entries $1$ on the diagonal with the rest of the row filled with the same value:
\begin{align*}
    &\vec{I} + \vec{D}^k \vec{M} = \\
    &\left(
    \begin{array}{c c c c}
         1 &  \frac{d_{12}^k(\widetilde{\alpha})}{n - \sum_{(r,s) \in \mathcal{V}}d_{rs}^k(\widetilde{\alpha}) + d_{12}^k(\widetilde{\alpha})} & \cdots & \frac{d_{12}^k(\widetilde{\alpha})}{n - \sum_{(r,s) \in \mathcal{V}}d_{rs}^k(\widetilde{\alpha}) + d_{12}^k(\widetilde{\alpha})} \\
         \frac{d_{13}^k(\widetilde{\alpha})}{n - \sum_{(r,s) \in \mathcal{V}}d_{rs}^k(\widetilde{\alpha}) + d_{13}^k(\widetilde{\alpha})} & 1 & \cdots & \frac{d_{13}^k(\widetilde{\alpha})}{n - \sum_{(r,s) \in \mathcal{V}}d_{rs}^k(\widetilde{\alpha}) + d_{13}^k(\widetilde{\alpha})} \\
         \vdots & \cdots & \ddots & \vdots \\
         \frac{d_{H, H-1}^k(\widetilde{\alpha})}{n - \sum_{(r,s) \in \mathcal{V}}d_{rs}^k(\widetilde{\alpha}) + d_{H, H-1}^k(\widetilde{\alpha})} & \frac{d_{H, H-1}^k(\widetilde{\alpha})}{n - \sum_{(r,s) \in \mathcal{V}}d_{rs}^k(\widetilde{\alpha}) + d_{H, H-1}^k(\widetilde{\alpha})} & \cdots & 1
    \end{array}
    \right).
\end{align*} Similarly, this matrix will be non-invertible whenever $n = \sum_{(r,s) \in \mathcal{V}}d_{rs}^k(\widetilde{\alpha})$, and it therefore is impossible to stay in any of the states in the interval $(\tau_{k-1}, \tau_k]$.

\end{document}